\documentclass[11pt]{article}
\usepackage{enumerate}
\usepackage{latexsym}
\usepackage{amsmath}
\usepackage{amssymb}
\usepackage{amsthm}
\usepackage{bbm}
\usepackage{epsfig}
\usepackage[latin1]{inputenc}
\usepackage{color}
\usepackage{multirow}
\usepackage{makecell}
\usepackage[table]{xcolor}

\usepackage{tikz}
\usetikzlibrary{shapes}
\usetikzlibrary{patterns}
\usetikzlibrary{arrows,positioning,decorations.pathmorphing,trees}

\usepackage{fullpage}

\newtheorem{theorem}{Theorem}[section]

\newtheorem{corollary}[theorem]{Corollary}

\newenvironment{restatetheorem}[1]
{\begingroup
  \def\thetheorem{\ref{#1}}
  \begin{theorem}}
{\end{theorem}
 \addtocounter{theorem}{-1}
 \endgroup}

\usepackage[noend]{algorithmic}
\usepackage{algorithm}
\floatname{algorithm}{Procedure}

\title{The Quarrelling Paradox Revisited:\\ A General Framework for a Class of Quarrels and Sabotage\\[.2cm]}
\author{A. Abizadeh\thanks{Department of Political Science, McGill University, Montreal, Canada, {\tt arash.abizadeh@mcgill.ca}; ORCID 0000-0002-8271-5194} 
\and A. Vetta\thanks{Department of Mathematics \& Statistics and School of Computer Science, McGill University, Montreal, Canada, {\tt adrian.vetta@mcgill.ca}}}

\begin{document}

\maketitle
\begin{abstract}
\noindent If a measure of voting power assigns players greater voting power because they no longer effectively cooperate,
then it displays the {\em quarrelling paradox} and violates the {\em quarrel postulate}.
However, we prove that certain types of quarrel increase some quarrellers' voting power on {\em any} proposed measure.
On the one hand, such quarrels are politically significant because they incentivize players to strategically join coalitions in order to sabotage them from within;
on the other, a postulate based on them cannot provide a reasonable normative criterion for evaluating measures of voting power.
We therefore formalize a general framework of quarrels -- comprising twelve conceptions distinguished according to symmetry, reciprocality, and strength --
and provide criteria for whether a conception provides a suitable basis for a reasonable quarrel postulate.
Although the two existing conceptions, proposed by Felsenthal and Machover and by Laruelle and Valenciano, do not, our framework's symmetric, weak conception does.

\end{abstract}

\newpage

\section{Introduction}\label{sec:intro}

It would be paradoxical if ruling out effective cooperation between two voters to achieve a jointly favoured outcome were
somehow to increase their voting power.
Following Kilgour (1974), the breakdown of effective cooperation between players has come to be known as a {\em quarrel} and,
following Brams (1975), a would-be increase in voting power due to quarrelling has come to be known as the {\em quarrelling paradox}.
A measure of voting power that displays the quarrelling paradox, in turn, can be said to violate a {\em quarrel postulate}.

Yet both the nature and status of the quarrelling paradox have been a matter of controversy and confusion.
Brams (1975: 181-82) himself concluded that two classic indices of voting power,
the Shapley-Shubik index (Shapley and Shubik 1954) and the normalized Banzhaf index (Banzhaf 1965, 1966), display the paradox, 
but this did not disturb his confidence in either.
Indeed, rather than counting against the indices, he thought the paradox revealed an important truth about voting power:
``Although one might suspect that they [the two quarrelling players] could only succeed in hurting each other, it is a curious fact that the quarrel...may 
actually redound to their benefit," such that ``there is an incentive for them to quarrel and increase their share of the voting power." 
Barry (1980: 193), by contrast, viewed Brams' conclusion as ``manifestly absurd";
he took the indices' apparent susceptibility to the quarrelling paradox obviously to disqualify them as measures of voting power.%
\footnote{For a more sympathetic yet still ambivalent view of Brams's conclusion, see Straffin (1982: 278-281).}

On the one hand, we agree with Barry that it would count against a proposed measure if it were to violate a suitably specified quarrel postulate.
But Barry's own criticisms of the Banzhaf and Shapley-Shubik indices miss the mark, for two reasons.
First, there is no reason to expect indices that measure a player's {\em relative share} to satisfy a quarrel postulate.
One's relative share of voting power depends not only on one's own power, but also on how much power others have;
a change detrimental to one player's voting power may be even more detrimental to others.
Thus two quarrelling players may hurt themselves, but hurt others (who depend on their cooperation, for example) even more (Felsenthal and Machover 1998: 240-41)
and so increase their relative share of voting power.
This is why, if players care about their power {\em relative} to others, they may indeed, as Brams thought, have reasons to instigate a quarrel.
Yet both Shapley-Shubik and Banzhaf are relative indices for which all players' scores sum to one.
The Banzhaf index, for example, is a normalization of  the Penrose-Banzhaf measure of voting power (Penrose 1946; cf. Felsenthal and Machover 1998, 2004).
As Felsenthal and Machover (1998: 40-41) note, by rescaling the Penrose-Banzhaf measure so that all players' scores sum to one,
the normalized Banzhaf index does not purport to measure -- as Barry supposed -- the amount of power a voter has;
it measures, rather, the player's ``relative share of total power."
So displaying a quarrel paradox would not count against either of these relative-share indices;
it could count only against absolute measures -- such as the Penrose-Banzhaf measure, from which the Banzhaf index is ultimately derivative.

Second, as we shall argue, previous formulations of the quarrel postulate have been based on conceptions of quarrelling that,
although reasonable as interpretations of what quarrelling is, are not fit to serve as the basis for a reasonable quarrel postulate.
We must keep track of two separate questions:
First, is a given {\em conception} of quarrelling a reasonable conception of the {\em concept} of quarrelling?%
\footnote{For the distinction between {\em conception} and {\em concept}, see, e.g., Rawls (1999: 5).}
And second, is the conception fit to serve as the basis for a reasonable quarrel postulate?
These questions are distinct: a conception of quarrelling may itself be reasonable,
but it may nevertheless be unreasonable to expect measures of voting power to satisfy a quarrel postulate constructed on its basis.
To serve as a normative standard, the quarrel postulate must be suitably specified.

On the other hand, we go beyond Brams and show that some types of quarrel present quarrellers with an opportunity to increase even their absolute voting power.
Indeed, in a key finding (Theorem \ref{thm:DNMQ}), we prove that certain types of quarrel would increase the voting power of some quarrellers on {\em any} measure of voting power.
The basic reason why some conceptions of quarrelling are unfit to serve as the basis for a reasonable quarrel postulate is that
they involve quarrels that reduce effective cooperation not only between the quarrelling players themselves,
but also for {\em other}, non-quarrelling players.
It turns out that these, more destructive types of quarrels -- including what we shall dub {\em strong} and {\em cataclysmic} quarrels -- 
do indeed provide a mechanism by which a quarrelling player could increase even its absolute voting power (and not merely, as Brams observed, its relative share).
A reasonable quarrel postulate must therefore be based on what we call a {\em weak} conception of quarrelling,
whose impact on effective cooperation is restricted to the quarrelling players themselves.

We therefore have two overarching goals in this paper.
The first is to specify a quarrel postulate that could justifiably serve as a normative criterion for evaluating proposed measures of voting power.
We do this by, first, articulating criteria by which to judge whether a given conception furnishes a reasonable interpretation of the concept of quarrelling;
second, identifying criteria by which to judge whether it would be reasonable to expect a measure of voting power to satisfy a quarrel postulate grounded in a given conception;
and third, specifying a conception of quarrelling, and a quarrel postulate based on it, that satisfy these criteria.
Our second overarching goal is to identify types of quarrelling that, albeit unsuitable to ground a reasonable quarrel postulate, provide opportunities for quarrellers to increase their voting power.
The former goal serves a normative or evaluative function, concerning measures of voting power, while the latter serves a descriptive one, concerning strategic interaction.

We can model the imposition of a quarrel in two ways  (Laruelle and Valenciano 2005a).
First, we could transform the original voting {\em structure},
imposing a constraint on the voting rule mapping each possible complete vote configuration or {\em division} onto outcomes.
Here we model quarrelling by reducing or wholly neutralizing the two players' ability effectively to cooperate in the 
divisions in which they vote together.
For example, take a simple voting game with two possible outcomes ({\sc yes} or {\sc no}),
where a {\sc yes}-outcome is guaranteed when at least two of three players $i$, $j$, and $k$ vote {\sc yes}.
We can model a quarrel between $i$ and $j$ by changing the mapping from divisions to outcomes
such that, once the quarrel breaks out, the votes of $i$ and $j$ are no longer sufficient to secure a {\sc yes}-outcome
in the divisions in which they vote {\sc yes} without $k$.

Second, we could model a quarrel by keeping the original voting structure intact but transforming the voting situation
by imposing constraints on the voting {\em behaviour} of players.
Here the mapping from divisions to outcomes remains the same;
we model quarrelling instead by imposing a constraint on the probability distribution of divisions.
For example, we reduce or set to zero the probability of divisions in which the two quarrelling players vote together (because they may refuse to do so).
So now we would say that, were $i$ and $j$ to vote {\sc yes} together without $k$, the outcome would still indeed be {\sc yes},
but the probability that the quarrelling players vote together is zero (or reduced).

The first approach yields a class of {\em structural} quarrel postulates, based on a structural model of quarrelling,
while the second approach yields a class of {\em behavioural} quarrel postulates, based on a behavioural model of quarrelling.
The latter approach, insofar as it models a quarrel as a disinclination on the part of the quarrelling players to vote together,
is perhaps more intuitive.
The problem, however, is that it is not suitable for evaluating the reasonability of so-called a priori measures of voting power (such as the Penrose-Banzhaf measure).
A priori voting power is one's voting power solely in virtue of the formal voting structure itself, that is, the sets of actors, their action profiles, 
and alternative outcomes on the agenda, as well as the decision function that maps divisions (i.e., combinations of actions) onto 
outcomes (Felsenthal and Machover 1998, 2003, 2004).
It therefore abstracts from the power one might have in virtue of the distribution of preferences (and consequent incentives for strategic interaction) 
within the voting structure -- an abstraction traditionally modelled by assuming equiprobable divisions.
The behavioural approach is unsuited to evaluating a priori measures because here the probability of each division is already fixed:
each division is equiprobable.
The class of behavioural quarrel postulates,
because based on a conception of quarrelling modelled by constraining division probabilities,
is suitable only for measures of voting power not restricted to equiprobable divisions.

Our focus here is on evaluating measures of voting power under the a priori assumption of equiprobable divisions.
We must therefore turn to the first approach, which models a quarrel as the inefficacy of co-action between the quarrelling players.
The question we address is therefore the correct formulation of the structural quarrel postulate that any reasonable measure of a priori 
voting power ought to satisfy.

Our thesis is that the two existing formulations of the structural quarrel postulate, by Felsenthal and Machover (1998) and 
Laruelle and Valenciano (2005a), respectively, face decisive shortcomings.
We accordingly propose a structural quarrel postulate for binary voting games,
based on a new conception of quarrelling, which is suitable for measures of a priori voting power and overcomes these defects.

The quarrel postulate concerns the comparison of a player's voting power in two voting games:
the initial voting game and a second game derived from it by inducing a quarrel.
Since we shall define this derivation or transformation only from initial voting games that are binary and monotonic,
we begin, in Section~\ref{sec:model}, by defining such games.
Then, in Section~\ref{sec:quarrel}, we identify what we take to be the essence of the concept of a quarrel and,
in its light, specify the set of necessary and sufficient conditions that any transformation function
should satisfy in order to count as inducing a quarrel.
We follow this, in Section~\ref{sec:postulate}, with the conditions that a conception of quarrelling should satisfy
to be fit to serve as the basis for a reasonable quarrel postulate.
In particular, we show why a reasonable quarrel postulate must be grounded in a conception of quarrelling that both preserves {\em monotonicity}
in the induced voting game (more precisely, it is not disposed to induce non-monotonicity over quarrellers) and is {\em symmetric}, i.e., carried out on both the {\sc yes}- and {\sc no}-voting sides.
We then show, in Sections~\ref{sec:FM} and~\ref{sec:LV}, that the two existing conceptions violate these conditions.
To identify our new conception and postulate,
and to contrast it with the two existing conceptions,
we formalize in Section~\ref{sec:framework} a general framework that distinguishes between three degrees of quarrelling (weak, strong, and cataclysmic),
between symmetric versus asymmetrical quarrels (Section~\ref{sec:three-degrees}),
and between {\em reciprocal} quarrels, in which both players quarrel with each other, and non-reciprocal ones, in which only one player quarrels with the other (Section~\ref{sec:nonrecip}).
This framework yields a typology of twelves quarrels
and is broad enough to encompass the conceptions of quarrelling formulated by both
Felsenthal and Machover and by Laruelle and Valenciano.
It also allows us to identify our new conception of quarrelling, which, unlike these two,
is indeed fit to serve as the basis for a reasonable quarrel postulate.
We therefore conclude (in Section \ref{sec:conclusion}) that the most general reasonable quarrel postulate
is based on a symmetric weak quarrel (whether reciprocal or not),
and show that the classic Shapley-Shubik index and Penrose-Banzhaf measure both satisfy this quarrel postulate
(despite failing to satisfy the Felsenthal-Machover and Laruelle-Valenciano postulates).
We also show that strong and cataclysmic quarrels can incentivize internal sabotage by furnishing opportunities for players to increase their voting power
by quarrelling with players whose voting coalition they strategically join.
Strikingly, players can sometimes increase their voting power even by engaging in {\em unilateral} or non-reciprocal cataclysmic quarrels.

\section{Voting Games}\label{sec:model}

Let $N=\{1,2,\dots, n\}$ be a finite set of players.
In a {\em binary voting game}, each player has two strategies (voting {\sc yes} or {\sc no}),
and there are two possible outcomes ({\sc yes} or {\sc no}).
In such a game, a division $\mathbb{S}=(S, N\setminus S)$ of the set $N$ is then an ordered partition of players
where the first element in the ordered pair is the set of {\sc yes}-voters
and the second element is the set of {\sc no}-voters in $\mathbb{S}$.
Thus, for $\mathbb{S}=(S, \bar{S})$, the subset $S\subseteq N$ comprises the set of {\sc yes}-voters 
and the subset $\bar{S}= N\setminus S$ comprises the set of {\sc no}-voters.
(Note the convention of representing a bipartitioned division by its first element in blackboard bold.)
We define $\mathcal{D}$ as the set of all possible divisions $\mathbb{S}$ of $N$.
The voting game then corresponds to a function $\mathcal{G}(\mathbb{S})$ mapping the set of all 
possible divisions $\mathcal{D}$ to the set of alternative outcomes $\mathcal{O}$=\{{\sc yes}, {\sc no}\}.

We say that any player whose vote corresponds to the division outcome is a {\em successful} player,
and that any division with a {\sc yes}-outcome is a {\em winning} division.
Let $\mathcal{W}$ be the collection of all sets of players $S$ such that $\mathcal{G}$($\mathbb{S}$)={\sc yes}, that is, if each 
member of $S$ were to vote {\sc yes}, they would be successful {\sc yes}-voters.
$\mathcal{W}$, the collection of {\em {\sc yes}-successful subsets} of $N$, commonly called {\em winning coalitions},
provides an alternative representation of the voting game $\mathcal{G}$.

Although we shall not restrict our definition of voting power to {\em monotonic} voting games,
we shall, for reasons that will become apparent,
define quarrels only over voting games that are initially monotonic (i.e., prior to the imposition of the quarrel):
\begin{quote}
\textbf{Monotonicity:} $\forall T \subseteq S \quad (T \in \mathcal{W} \ \Rightarrow\ S \in \mathcal{W})$. 
\end{quote}
Monotonicity states that if a division outcome is {\sc yes}, then the outcome of any division in which at least 
the same players vote {\sc yes} will also be {\sc yes}.
The literature on voting power also typically assumes that voting games satisfy {\em non-triviality}:
\begin{quote}
\textbf{Non-Triviality:} $\exists S |$ $S \in \mathcal{W}$ and $\exists S |$ $S \notin \mathcal{W}$.
\end{quote}
Non-triviality states that not all divisions yield the same outcome.
Non-triviality is typically assumed because a voting game that fails to satisfy it is of little interest for measures of voting power:
such a game would effectively be a structure in which voters never make a difference to the outcome and
therefore automatically one in which voters have no voting power at all. (It is effectively not a voting game.)
A monotonic binary voting game that satisfies non-triviality is called a {\em simple voting game}.
Monotonicity and non-triviality together ensure that simple voting games also satisfy {\em unanimity}:
\begin{quote}
\textbf{Unanimity:} $N\in \mathcal{W}$ and $\emptyset \notin \mathcal{W}$.
\end{quote}

\section{What is a Reasonable Conception of a Quarrel?}\label{sec:quarrel}

We seek to define a class of quarrels and quarrel postulates for quarrels imposed on initially monotonic binary voting games.
Our task is to define a conception of quarrelling that is itself reasonable, on the one hand, and
fit to serve as the basis for a reasonable quarrel postulate, on the other.
A {\em conception} of quarrelling is reasonable insofar as it captures core intuitions about 
what the concept of a quarrel is.
A quarrel {\em postulate} is reasonable in the sense that it would be reasonable to expect a measure of voting power to satisfy it.
In this section, we analyze the concept of a quarrel,
and specify the core property we assume a reasonable conception should satisfy.
We turn to the quarrel postulate in the next section.

\subsection{Reducing Effective Cooperation}\label{sec:EC}

We understand a quarrel between two players to be a phenomenon in which their {\em effective cooperation} is reduced.
Since we assume that quarrels are imposed on an initially motonotonic binary game --
which renders sincere voting a dominant strategy and hence rules out strategic voting --
we use the term {\em cooperation} to refer to a circumstance in which two players agree in their votes,
that is, they vote on the same side, {\sc yes} or {\sc no}.
Recall that the behavioural approach models a quarrel by reducing the likelihood the quarrelling players agree in their votes.
The structural approach we pursue here, however, models a quarrel by reducing or neutralizing the effectiveness of their voting together on the same side.
A quarrel is therefore operative under this approach only when their votes do indeed agree;
the quarrel consists of the fact that, when the players vote on the same side,
their votes do not work together as effectively to contribute to the outcome for which they vote.
Thus the difference between the two approaches consists in how each models the reduction
or neutralization of effective cooperation that constitutes quarrelling:
whereas the behavioural approach reduces {\em cooperation},
the structural approach reduces {\em effectiveness}.

Two players' cooperation is effective in our sense if they are each successful and contribute to the outcome.
We assume that one way in which a player may contribute is by casting a decisive vote.
A player (or its vote) is {\em decisive} in a division if, holding all other votes constant,
it could have unilaterally changed the outcome if it had voted differently.
We say a player is {\sc yes}-decisive if it votes {\sc yes} and is decisive,
and is {\sc no}-decisive if it votes {\sc no} and is decisive.
For binary monotonic voting games, this can be formalized as follows:
player $i$ is {\sc yes}-decisive in $\mathbb{S}$
if and only if $i\in S\in \mathcal{W}$ but $S\setminus\{i\}\notin \mathcal{W}$,
is {\sc no}-decisive if and only if $i\notin S\notin \mathcal{W}$ but $S\cup\{i\}\in \mathcal{W}$,
and is decisive if and only if it is either {\sc yes}- or {\sc no}-decisive.
Note, however, that for non-monotonic games, we must add
that $i$ is {\sc yes}-decisive in $\mathbb{S}$ also if $i\in S \notin \mathcal{W}$ but $S\setminus\{i\} \in \mathcal{W}$,
and that $i$ is {\sc no}-decisive in $\mathbb{S}$ also if $i\notin S \in \mathcal{W}$ but $S\cup \{i\} \notin \mathcal{W}$.
In other words, in non-monotonic games a player's vote may be decisive even when it {\em disagrees} with the outcome.
Being decisive in this way is consistent with the notion of having and exercising voting power:
it involves exercising voting power to satisfy one's preferences by voting {\em strategically}.
In non-monotonic -- and, indeed, in non-binary -- games, rational actors may
have a reason to vote strategically {\em against} the outcome they prefer.

In order to leave open the possibility that players may contribute and hence effectively cooperate even in causally overdetermined divisions,
however, we do not assume that players must be decisive in order to contribute.
Being decisive is sufficient but not necessary.
We assume instead that a player cannot contribute to outcomes in $\mathcal{G}$ if it is a dummy, i.e., not ever decisive in any division;
concomitantly, we assume that two players effectively cooperate in a division in $\mathcal{G}$ only if they are each decisive in at least some division in $\mathcal{G}$.

Since the structural approach to modelling quarrelling is less intuitive than the behavioural one,
we here motivate the former approach, necessary for evaluating a priori power,
by presenting a model that furnishes an intuitive interpretation.
We can model structural conceptions by disaggregating {\em cooperation} and {\em effectiveness}
into two distinct moments or stages in the voting procedure.
In stage~1, the players vote; to vote on the same side is to cooperate.
In stage~2, the players on the same side must then jointly present their side's votes to the tabulator.
This second stage is when a quarrel may break out (between players who, in stage 1, voted on the same side),
with the result that those affected by the quarrel are unable or refuse to present their vote to the tabulator.
It is only by presenting their votes together to the tabulator in the second stage that first-stage cooperation can become effective.
The model works by incorporating the quarrel at stage~2,
thereby allowing us to maintain the a prioristic approach in stage 1, to which the quarrel postulate applies.%
\footnote{This obviates the doubts raised by Felsenthal and Machover (1998: 238)
about the reasonability of a structural quarrel postulate for a priori measures of voting power on the grounds
that ``the concept of {\em quarrelling} oversteps the limits of the a prioristic terrain.''}

The advantage of the structural model is that it allows us not only to develop a quarrel postulate for a priori power, but also,
as we shall see, to model the phenomenon of internal {\em sabotage}, whereby a player may strategically vote for its dispreferred outcome,
ostensibly cooperating with others to bring it about, only to quarrel with them in order to render their cooperation ineffective.

\subsection{Three Quarrel Properties}\label{sec:3QP}

To evaluate the reasonability of any (structural) conception of a quarrel,
we begin by first identifying the core property that we assume any reasonable (structural) conception of a quarrel should possess,
namely {\em cooperative-success-reduction};
an interpretation of that property, which we call {\em internal motivation};
and two further properties relevant for distinguishing the type of quarrel it is, namely, {\em symmetry} and {\em reciprocality}.
We begin with an informal characterization of each property for intuition,
and then formalize them.

We assume that the essence of a quarrel can be articulated by the property we label cooperative-success-reduction (CSR).
We assume, in other words, that any conception must satisfy this property to be a reasonable conception of a quarrel,
and that any conception satisfying it is ipso facto a reasonable conception.
CSR captures two core notions:
first, that quarrelling would never benefit the effectiveness of the cooperation of voters whose votes agree (even if it might, as we shall see, confer other, strategic benefits);
and second, that quarrelling can sometimes be harmful in this respect.
CSR therefore holds that when the votes of two players $i$ and~$j$ agree,
they are never {\em more successful} in securing the outcome for which they both vote if they quarrel than if they do not,
and are sometimes {\em less successful}.
We assume CSR captures the essence of a quarrel because we take the notion of quarrelling to refer to diminishing effective cooperation,
and take the notion of effective cooperation to imply helping each other to secure outcomes.%
\footnote{Compare CSR with a weaker property, that quarrelling players are not successful in more divisions {\em overall}.
This latter property (which is implied by CSR, but not vice versa) permits quarrelling to {\em improve} success in some divisions
as long as such improvement is offset by success-reduction in other divisions.
This is too weak to capture the notion of quarrelling precisely because it implies
that in some circumstances quarrelling might improve the effectiveness of two players' cooperation.}
Thus, intuitively, when effective cooperation between two players diminishes,
their effectiveness in helping each other to secure outcomes for which they both vote never improves and sometimes is lowered.%
\footnote{We say {\em cooperative}-success-reduction because CSR concerns players' success in divisions in which their votes {\em agree}.
This contrasts with players' reduced success in divisions in which their votes {\em disagree},
which corresponds to the concept of an {\em ambush}, not quarrel.
We shall return to the ambush concept in Section~\ref{sec:framework}.}

We also assume that any conception of quarrelling must be compatible with interpreting the reduction or elimination of cooperative success
as {\em internally motivated}, in the sense that the reason for reduction refers to a quarrelling player's own agency.
By contrast, consider a scenario in which the cooperative success of two unwavering friends is reduced
because a third party interferes with the effectiveness of their cooperation:
this scenario could also be formalized by CSR,
but it does not correspond to the intuitive notion of a quarrel between the two players.
It corresponds, rather, to the concept of what we might call an {\em external attack}.

For the second property, symmetry, consider the case in which, as a result of their quarrel,
two members of the legislature no longer effectively cooperate to {\em pass} proposed legislation,
but remain willing and able to cooperate effectively to {\em block} legislation,
thus quarrelling only on the {\sc yes} side.
This would be an asymmetric {\sc yes}-quarrel.
For intuition, imagine that the reason they quarrel is that each suspects the other's enmity,
and so doubts the advisability of the outcome for which they vote when their enemy also votes in favour.
An asymmetric {\sc yes}-quarrel might then arise because each player doubts the advisability only of {\sc yes}-outcomes when the other also votes in favour.%
\footnote{Other interpretations of an asymmetric quarrel are also available.
Imagine the players suspect each other,
but are extremely risk-averse when it comes to change and hence will accept anyone's cooperation on the {\sc no}-side when they vote for the status quo,
but not so on the {\sc yes}-side, and are therefore quick to give in to their suspicions over any changes to the status quo.}
Thus we informally define a {\em symmetric} quarrel as one equally carried out on both the {\sc yes} and {\sc no} sides.

Finally, consider the third property, namely reciprocality.
Take the case in which player $i$ has a quarrel with $j$, but not vice versa.
Here $i$ does not effectively cooperate with $j$,
even though~$j$ is willing and able effectively to cooperate with $i$ if its cooperation is effectively reciprocated.
There are two ways to interpret this.
First, it might be that $i$ {\em cannot} effectively cooperate with $j$ because someone {\em else} (whether $j$ or not) has neutralized $i$'s capacity to do so.
The trouble with this interpretation is that it does not correspond to the idea that $i$ has a {\em quarrel} with $j$:
recall that a quarrel intrinsically depends on the agency of the quarrelling player itself.
This interpretation corresponds, rather, to the idea that someone other than $i$ paralyzes $i$'s effective cooperation with~$j$.
Second, it might be that $i$ {\em does not} effectively cooperate with $j$ because $i$ itself is unwilling (or undermines its own capacity) to do so.
This corresponds to the intuitive concept of a quarrel as we construe it.
But if $j$ has no quarrel with $i$, and so is willing and able effectively to cooperate with $j$,
then $i$'s quarrel with $j$ is a {\em non-reciprocal} quarrel of $i$ against $j$.
For intuition, imagine that $i$ suspects or cannot bear $j$ and hence does not effectively cooperate with $j$,
but that $j$ is entirely indisposed to quarrelling of any kind:
other players' quarrels simply do not affect $j$'s willingness and capacity to function with others if they too are willing and able.
Thus we informally define a {\em reciprocal} quarrel as one in which both players equally quarrel with each other.

We now turn to providing formal statements for each of these three properties.

\subsection{Cooperative-Success-Reduction (CSR)}\label{sec:CSR}

Let $\mathcal{G}$ be a monotonic binary voting game,
$\mathbb{G}$ be the set of all monotonic binary voting games,
$\hat{\mathcal{G}}$ be a voting game derived from $\mathcal{G}$ by imposing a quarrel between two of $\mathcal{G}$'s players, $i$ and $j$,
and $\mathcal{Q} : \mathcal{G} \rightarrow \hat{\mathcal{G}}$ be the {\em transformation rule} or function
that maps each division outcome in $\mathcal{G}$ to the corresponding division outcome in $\hat{\mathcal{G}}$.
The transformation rule $\mathcal{Q}$, by which the voting game $\hat{\mathcal{G}}$ is derived from $\mathcal{G}$,
can be thought of as formalizing a particular conception of what a quarrel is,
that is, a formalization of the conception imposed on $\mathcal{G}$ and reflected in $\hat{\mathcal{G}}$.
We formulate CSR, which characterizes the transformation rule $\mathcal{Q}$,
in terms of the {\sc yes}-successful voter subsets $\mathcal{W}$ in $\mathcal{G}$
and $\hat{\mathcal{W}}$ in $\hat{\mathcal{G}}$:
\begin{quote}
{\bf Cooperative-Success-Reduction (CSR).}
For any $S$ containing neither $i$ nor $j$: 
\begin{align*}
& \text{{\bf On the {\sc Yes} Side}} \tag{{\bf YQ}} \label{eq:YQ}\\
&\ \forall \mathcal{G}\in \mathbb{G}:\ & S\cup \{i,j\} \in \hat{\mathcal{W}} & \implies  S\cup \{i,j\} \in \mathcal{W} \tag{YQ-1} \label{eq:YQ1}\\
&\ \exists \mathcal{G}\in \mathbb{G}:\ & \exists S \mid \ S\cup \{i,j\} \notin \hat{\mathcal{W}} & \ \ \wedge\ \ \ S\cup \{i,j\} \in \mathcal{W} \tag{YQ-2} \label{eq:YQ2}\\[0.2cm]
& \text{{\bf On the {\sc No} Side}} \tag{{\bf NQ}} \label{eq:NQ}\\
&\ \forall \mathcal{G}\in \mathbb{G}:\ & S\in \hat{\mathcal{W}} &\Longleftarrow S\in \mathcal{W} \tag{NQ-1} \label{eq:NQ1}\\
&\ \exists \mathcal{G}\in \mathbb{G}:\ &\exists S \mid \ S\in \hat{\mathcal{W}} &  \ \ \wedge\  S\notin \mathcal{W} \tag{NQ-2} \label{eq:NQ2}
\end{align*}
\end{quote}
Here $\wedge$ is the logical operator {\sc and}. Observe that CSR has two parts.
The first, YQ, formalizes the idea that $i$ and $j$ have a quarrel on the {\sc yes} side.
\ref{eq:YQ1} states that if $i$ and $j$ vote {\sc yes} in a division in which the outcome is {\sc no} in the original game $\mathcal{G}$,
then the outcome of that division will also be {\sc no} once they quarrel in $\hat{\mathcal{G}}$.
This implies that $i$ and $j$ are {\em never more} {\sc yes}-successful thanks to their quarrel when their votes agree on the {\sc yes} side,
which means their quarrel never renders cooperation between them more effective on the {\sc yes} side.
\ref{eq:YQ2} states that there exists at least one monotonic binary game $\mathcal{G}$ with at least one division in which $i$ and $j$ vote {\sc yes}
and whose outcome is {\sc yes} in the original game $\mathcal{G}$ but {\sc no} once they quarrel in $\hat{\mathcal{G}}$.
This implies that, for such a game, in at least one division in which both players vote {\sc yes} they are {\em less} {\sc yes}-successful due to their quarrel,
which means their quarrel renders cooperation between them less effective on the {\sc yes} side in at least one division.

The second part, NQ, formalizes the idea that $i$ and $j$ have a {\sc no}-quarrel.
\ref{eq:NQ1} states that if $i$ and $j$ vote {\sc no} in a division in which the outcome is {\sc yes} in the original game $\mathcal{G}$,
then the outcome of that division will also be {\sc yes} once they quarrel in $\hat{\mathcal{G}}$.
This implies that $i$ and $j$ are {\em never more} {\sc no}-successful thanks to their quarrel,
which means their quarrel never renders cooperation between them more effective on the {\sc no} side.
\ref{eq:NQ2} states that there exists at least one monotonic binary game $\mathcal{G}$ with at least one division in which $i$ and $j$ vote {\sc no}
and whose outcome is {\sc no} in the original game $\mathcal{G}$ but {\sc yes} once they quarrel in $\hat{\mathcal{G}}$.
This implies that, for such a game, in at least one division in which both players vote {\sc no} they are {\em less} {\sc no}-successful due to their quarrel,
which means their quarrel renders cooperation between them less effective on the {\sc no} side in at least one division.

Note that neither \ref{eq:YQ2} nor \ref{eq:NQ2} requires that the transformation reduce the quarrelling players' success 
for {\em all} pairs $\mathcal{G}$ and $\hat{\mathcal{G}}$.
This is because there may be no effective cooperation between them in the first place in the initial game $\mathcal{G}$;
if so, then there would be nothing between them for a quarrel to reduce.
Consider, for example, a monotonic binary voting game $\mathcal{G}$ with three players in which the first player is a dictator,
such that $\mathcal{W}=\{\{1\}, \{1,2\}, \{1,3\}, \{1,2,3\}\}$.
In this structure, because player 1 {\em unilaterally} dictates the outcome and the others are dummies,
there can be no effective cooperation between it and them in any case;
thus imposing a quarrel between players 1 and 2 need not induce any change whatsoever in this voting game.
But a transformation rule satisfying \ref{eq:YQ2} or \ref{eq:NQ2} will incur a change for other voting games.

We say that a conception of a quarrel satisfying YQ (=\ref{eq:YQ1}$\wedge$\ref{eq:YQ2}) satisfies CSR on the {\sc yes} side;
a conception satisfying NQ (=\ref{eq:NQ1}$\wedge$\ref{eq:NQ2}) satisfies CSR on the {\sc no} side;
and a conception satisfying both YQ and NQ satisfies CSR on both sides.
On our account, any reasonable conception of a {\sc yes}-quarrel must satisfy YQ;
any reasonable conception of a {\sc no}-quarrel must satisfy NQ;
and any reasonable conception of a symmetric quarrel must satisfy CSR on both sides.
This articulates what we assume is the core intuitive idea of quarrelling, namely,
that a group containing a quarrelling couple will never be more successful and sometimes be less successful in cooperatively securing outcomes than it would 
have been had the couple not been quarrelling.

Note, finally, that neither \ref{eq:YQ2} nor \ref{eq:NQ2} requires that the transformation wholly neutralize or {\em eliminate}
effective cooperation between the quarrelling players in $\mathcal{G}$: it merely requires reduction.
Quarrels that eliminate effective cooperation between the quarrelling players would satisfy a stronger variant of each condition,
namely, that for any $S$ containing neither $i$ nor $j$:
\begin{align*}
\forall \mathcal{G} & \in \mathbb{G}:\  & S\cup \{i,j\} \notin \hat{\mathcal{W}} & \ \Longleftarrow\  S\cup \{i,j\} \in \mathcal{W} 
	\ \wedge\  S\cup \{i\} \notin \mathcal{W} \ \wedge\  S\cup \{j\} \notin \mathcal{W} \tag{YQ'-2} \label{eq:YQ'2}\\[0.2cm]
\forall \mathcal{G} & \in \mathbb{G}:\  &  S \in \hat{\mathcal{W}} & \ \Longleftarrow\  S \notin \mathcal{W} 
	\ \wedge\ S\cup \{i\} \in \mathcal{W} \ \wedge\ S\cup \{j\} \in \mathcal{W} \tag{NQ'-2} \label{eq:NQ'2}
\end{align*}

\noindent \ref{eq:YQ'2} and \ref{eq:NQ'2} say that whenever, in a division in $\mathcal{G}$, both $i$ and $j$ are decisive,
then in $\hat{\mathcal{G}}$ they are both unsuccessful in that division.
This means that their cooperation is wholly ineffective in $\hat{\mathcal{G}}$ for any division.%
\footnote{Recall that we left open the possibility of effective cooperation in causally overdetermined divisions in which both players
are not fully decisive (but in which one or both may be partially efficacious).
But we also assumed that partial efficacy in one division (in which the players' votes agree) is parasitic on being decisive in at least one division
(in which the players' votes agree) in the game.}
Satisfying \ref{eq:YQ'2} and \ref{eq:NQ'2} obviously implies satisfying \ref{eq:YQ2} and \ref{eq:NQ2}, respectively.
(All the conceptions of quarrelling we examine below satisfy \ref{eq:YQ'2} and/or \ref{eq:NQ'2}
because they eliminate rather than merely reduce effective cooperation between the quarrelling players.)

\subsection{Symmetry}\label{sec:symmetry}
To formalize our second property, {\em symmetry},
we define a game $\mathcal{G}^C$ to be the {\em complement} of $\mathcal{G}$ if, for any $S\subseteq N$,
$$S\in \mathcal{W} \iff \bar{S}\notin \mathcal{W}^C$$ 
Equivalently, $S$ is {\sc no}-successful in $\mathcal{G}^C$ if and only if it is {\sc yes}-successful in $\mathcal{G}$.
Thus the complement $\mathcal{G}^C$ would yield an identical game to $\mathcal{G}$ if the {\sc yes/no} vote labels were interchanged.

It follows that for a quarrel to be symmetric with respect to {\sc yes/no},
the games we obtain by imposing a quarrel on $\mathcal{G}$ and on $\mathcal{G}^C$ must themselves be complements!
Any violation of this property implies that inducing the quarrel has a different effect when we swap the labels of the votes:

\begin{quote}
{\bf Symmetry:} For any game $\mathcal{G}\in \mathbb{G}$, it holds that $(\hat{\mathcal{G}})^C = \hat{(\mathcal{G}^C)}$.
\end{quote}

An {\em asymmetric} quarrel, in turn, is any quarrel that is not symmetric.
This includes {\em purely} asymmetric quarrels, in which the players quarrel on one side only ({\sc yes} or {\sc no}),
and {\em quasi}-symmetric quarrels, in which the players quarrel on both sides,
but their quarrel on one side is not identical to the other.
(We shall further formalize this once we introduce our framework).
Since in this paper we treat only the pure case, we use ``asymmetric'' to refer to purely asymmetric quarrels.

\subsection{Reciprocality}\label{sec:reciprocality}
To formalize our third property, {\em reciprocality},
let $\hat{\mathcal{G}}^{i,j}$ represent the game induced by a quarrel of~$i$ with $j$
and $\hat{\mathcal{G}}^{j,i}$ represent the game induced by a quarrel of~$j$ with $i$.
The reciprocality of quarrel $\mathcal{Q}$ simply means that the induced game $\hat{\mathcal{G}}^{i,j}$ is identical to the induced game $\hat{\mathcal{G}}^{j,i}$:
\begin{quote}
{\bf Reciprocality:} For any game $\mathcal{G}\in \mathbb{G}$, it holds that $\hat{\mathcal{G}} 
	= \hat{\mathcal{G}}^{i,j} = \hat{\mathcal{G}}^{j,i}$.
\end{quote}

A {\em non-reciprocal} quarrel, in turn, we define as any quarrel that is not reciprocal.
This includes {\em purely} non-reciprocal quarrels, in which one player quarrels against another who has no quarrel with it,
and {\em quasi}-reciprocal quarrels, in which both players quarrel with each other, but not equally.
(Again, we shall further formalize this once we introduce our framework).
Since in this paper we treat only the pure case, we use ``non-reciprocal'' to refer to purely non-reciprocal quarrels.

\subsection{Reasonable Conceptions of Quarrelling}\label{sec:reasonable}

As far as {\em quarrelling} itself is concerned, any conception $\mathcal{Q}$ that satisfies CSR (on at least one side)
counts as a reasonable formal conception -- whether or not the conception satisfies symmetry or reciprocality.
What symmetry and reciprocality do is help specify the quarrel's type;
we find nothing unreasonable per se about an asymmetric or non-reciprocal conception of quarrelling.
One could even combine symmetry and reciprocality in various ways:
a quarrel could be reciprocal on the {\sc yes} side but not on the {\sc no} side,
or $i$'s quarrel with $j$ could be symmetric but $j$'s quarrel with $i$ asymmetric.
We leave aside these hybrid types for the sake of brevity.
CSR leaves unspecified just how deleterious the quarrel is,
and so {\em how much} less successful the group to which the quarrelling couple belongs would be.
We specify this magnitude in Section~\ref{sec:three-degrees} where we present a class of quarrels based on three degrees of quarrelling.

\section{What is a Reasonable Quarrel Postulate?}\label{sec:postulate}

To define the quarrel postulate and corresponding paradox, in addition to the concept of a quarrel
we require that of voting power.
Although we have defined quarrels only over initially monotonic binary voting games,
we define a {\em measure of voting power} more broadly for binary voting games in general,
as a function $\Psi$ that assigns to each player $i$ a nonnegative real number $\Psi_i \geq 0$
and that satisfies two sets of basic adequacy postulates concerning voting power under a prioristic assumptions
(which we represent with the lower case $\psi$):
the {\em iso-invariance} postulate, according to which the a priori voting power $\psi_i$ of any player $i$ remains the same between two isomorphic games;
and the {\em dummy} postulates, according to which a player has zero a priori voting power if and only if it is a dummy
(i.e., is not decisive in any division),
and the addition of a dummy to a voting structure leaves other players' a priori voting power unchanged (Felsenthal and Machover 1998: 222).
We can now define the {\em quarrel postulate} in its standard form on the basis of a quarrel in $\hat{\mathcal{G}}$
between $i$ and $j$ derived, via $\mathcal{Q}$, from a monotonic binary voting game $\mathcal{G}$
(where $\hat{\psi}$ represents players' a priori voting power in the derived game $\hat{\mathcal{G}}$):

\begin{quote}
{\bf (Standard) Quarrel Postulate:} A measure of voting power $\Psi$ satisfies the (standard) quarrel postulate 
based on $\mathcal{Q}$ if and only if $\hat{\psi}_ i\le \psi_i$ and $\hat{\psi}_ j\le \psi_j$ for any $\mathcal{G}\in \mathbb{G}$.
\end{quote}

\noindent A measure of voting power displays the {\em quarrelling paradox} if and only if it violates the quarrel postulate.
Thus, a violation arises if the a priori voting power of at least one quarrelling pair increases after quarrelling.

Just as it is possible to disaggregate voting power over binary games into its two components, namely, {\sc yes}-voting power and {\sc no}-voting power,
it is also possible to disaggregate the quarrel postulate into two corresponding parts.
We define {\sc yes}-voting power as that part of a player's voting power based solely on its potential {\sc yes}-votes,
and {\sc no}-voting power as the part based solely on its potential {\sc no}-votes.
We therefore define a {\em measure of {\sc yes}-voting power} for binary voting games as a function $\Psi^+$
that assigns to each player $i$ a nonnegative real number $\Psi^+_i \geq 0$,
satisfies the basic adequacy postulates,
and is a function of only those divisions in which $i$ votes {\sc yes}.
Symmetrically, one can define a measure of {\sc no}-voting power $\Psi^-$ such that $\Psi_i = \Psi^+_i + \Psi^-$.
We can now define the corresponding components of the standard quarrel postulate:

\begin{quote}
{\bf {\sc Yes}-Voting-Power Quarrel Postulate:} A measure of voting power $\Psi$ satisfies the {\sc yes}-voting-power 
quarrel postulate based on $\mathcal{Q}$ if and 
only if $\hat{\psi}_i^+\le \psi_i^+$ and $\hat{\psi}_ j^+\le \psi_j^+$ for any $\mathcal{G}\in \mathbb{G}$.\\[-0.3cm]

{\bf {\sc No}-Voting-Power Quarrel Postulate:} A measure of voting power $\Psi$ satisfies the {\sc no}-voting-power 
quarrel postulate based on $\mathcal{Q}$ if and 
only if $\hat{\psi}_i^-\le \psi_i^-$ and $\hat{\psi}_ j^-\le \psi_j^-$ for any $\mathcal{G}\in \mathbb{G}$.
\end{quote}

\noindent The satisfaction of both these non-standard postulates of course implies the satisfaction of the standard postulate.

\subsection{Transformation Monotonicity}\label{sec:monotonicity}

Whether or not a quarrel postulate is reasonable will depend on the conception of quarrelling on which it is based.
The conception itself must of course be reasonable, that is, it must satisfy CSR.
Although we define quarrels only over initially monotonic games,
CSR permits the derived quarrelling game to be non-monotonic.
(Note that we have defined voting power over non-monotonic voting games as well.)
When a transformation rule $\mathcal{Q}$ preserves or fails to preserve monotonicity in the induced game $\hat{\mathcal{G}}$,
we shall say that the transformation rule $\mathcal{Q}$ itself is monotonic or non-monotonic, respectively:

\begin{quote}
{\bf Transformation Monotonicity (TM):} \qquad \quad $\forall \mathcal{G} \in \mathbb{G}: \forall T \subseteq S \ (T \in \hat{\mathcal{W}} \ \Rightarrow\ S \in \hat{\mathcal{W}})$.\\
{\bf Transformation Non-Monotonicity (TNM):} \ $\exists \mathcal{G} \in \mathbb{G}:  \ \exists T \subseteq S \ (T \in \hat{\mathcal{W}} \ \wedge\ S \notin \hat{\mathcal{W}})$.
\end{quote}

\noindent We think that CSR's compatibility with non-monotonic quarrels is justified:
we find nothing unreasonable about a conception of quarrelling that induces non-monotonic voting games once the players quarrel.

However, as we shall now argue, a conception of quarrelling that is disposed to induce non-monotonicity over quarrellers
is not fit to serve as the basis of a reasonable {\em quarrel postulate}.
To be reasonable, a quarrel postulate must be based on a conception of quarrelling that, beginning with any initially monotonic game ${\mathcal{G}}$,
is not disposed to induce non-monotonicity over the quarrellers in the quarrelling game $\hat{\mathcal{G}}$.
Monotonicity properties are significant here because if $\hat{\mathcal{G}}$ departs from monotonicity in a way that involves the quarrelling players,
then it would not be reasonable to expect a measure of voting power to avoid succumbing to a quarrelling paradox,
because the paradox, rather than reflecting any defect in the measure itself,
may merely be an artifact of the voting structure's departure from monotonicity.
Intuitively, there is no reason to expect a measure of voting power to satisfy the quarrel postulate over games that are non-monotonic over the quarrellers,
because just as a set of players may, by definition, be more {\em successful} in a non-monotonic game if some of their members vote against them
(not cooperating),
so too might some players become more {\em powerful} if they were subject to a quarrel (not cooperating effectively).

We can formalize this intuition as follows.
Recall that monotonicity is violated in $\hat{\mathcal{G}}$ if there exists a {\em violating pair} of divisions $\{\mathbb{T},\mathbb{S}\}$,
such that $T \subseteq S$ and $T \in \hat{\mathcal{W}}$ but $S \notin \hat{\mathcal{W}}$.
We say that the non-monotonicity in $\hat{\mathcal{G}}$ is {\em over quarrellers} if the two divisions in the violating pair are identical except that the quarrelling duo $\{i,j\}$ agree in one and disagree in the other, i.e., player $j$'s coalition is successful without $i$'s contribution, either on the {\sc yes} or {\sc no} side, but loses when $i$ joins it, or vice versa:
\begin{quote}
\textbf{Non-Monotonicity over Quarrellers (NMQ):} 
The induced game $\hat{\mathcal{G}}$ is non-monotonic over quarrellers $\{i,j\}$ iff:
\begin{align*}
\exists S\ (i,j\notin S):\ \exists T,X \in \mathcal{Z}(S):\ (T \subseteq X \ \wedge\ T \in \hat{\mathcal{W}} \ \wedge\ X \notin \hat{\mathcal{W}}),
\end{align*}
where $\mathcal{Z}(S)=\{S,\,S\cup\{i\},\,S\cup\{j\},\,S\cup\{i,j\}\}$.
\end{quote}

\noindent Concomitantly, we say that a quarrel $\mathcal{Q}$ is {\em transformation non-monotonic over quarrellers} if it induces at least one such game $\hat{\mathcal{G}}$:

\begin{quote}
\textbf{Transformation Non-Monotonicity over Quarrellers (TNMQ):}
The quarrel $\mathcal{Q}$ is transformation non-monotonic over quarrellers $\{i,j\}$ iff:
\begin{align*}
\exists \mathcal{G}\in\mathbb{G}:\ \exists S\ (i,j\notin S):\ \exists T, X \in\mathcal{Z}(S):\ (T\subseteq X\ \wedge\ T\in\hat{\mathcal{W}}\ \wedge\ X\notin\hat{\mathcal{W}}),
\end{align*}
where $\mathcal{Z}(S)=\{S,\ S\cup\{i\},\ S\cup\{j\},\ S\cup\{i,j\}\}$. \\
\end{quote}

We now specify a property consisting in a tendency to induce non-monotonicity over quarrellers when the original game has a certain vulnerability, namely,
it contains at least one division in which, prior to the quarrel in $\mathcal{G}$, the quarrelling pair votes together and is successful, and one of them is decisive but the other is not.
We say that a conception $\mathcal{Q}$ is {\em disposed to induce non-montonicity over quarrellers}
if, for any quarrelling pair $\{i,j\}$ and for any $\mathcal{G}$ subject to this vulnerability (condition $V(S)$),
the transformation rule $\mathcal{Q}$ would induce non-monotonicity over at least one such division by rendering $i$ and $j$ unsuccessful in it when they quarrel in $\hat{\mathcal{G}}$ (condition $I(S)$).

\begin{quote}
\textbf{Disposition to Induce Non-Monotonicity Over Quarrellers (DNMQ):}
The transformation rule $\mathcal{Q}$ is disposed to induce non-monotonicity over quarrellers iff, 
for any pair $\{i,j\}$ where $i$ quarrels with $j$:
\begin{align*}
\forall \mathcal{G} \in \mathbb{G} : \qquad & V(S) \ \Longrightarrow \ I(S)
\end{align*}
\qquad where:
\begin{align*}
V(S) &:= \ \exists S \ (i,j \notin S) \ \mid \ S \ \cup \{i\} \in \mathcal{W} \ \wedge \ S \cup \{j\} \notin \mathcal{W},\\
I(S) &:= \ \exists S \ (i,j \notin S) \ \mid \ (S \cup \{i\} \in \mathcal{W} \cap \hat{\mathcal{W}} \ \wedge \ S \cup \{i,j\} \notin \hat{\mathcal{W}}) \ \vee \ (S \cup \{j\} \notin \mathcal{W} \cup \hat{\mathcal{W}} \ \wedge \ S \in \hat{\mathcal{W}}).
\end{align*}

\end{quote}

\noindent Here $\vee$ is the logical operator {\sc or}.
Notice that since $\mathcal{G}$ is monotonic, if $S \cup \{i\} \in \mathcal{W}$ then $S \cup \{i,j\} \in \mathcal{W}$ and if $S \cup \{j\} \notin \mathcal{W}$ then $S \notin \mathcal{W}$ in $V(S)$.
When $I(S)$ holds because of the first parenthetical expression prior to $\vee$, $\mathcal{Q}$ is disposed to induce non-monotonicity over quarrellers on the {\sc yes} side;
when $I(S)$ holds because of the second parenthetical expression, $\mathcal{Q}$ is disposed to induce non-monotonicity over quarrellers on the {\sc no} side.%
\footnote{A stronger variant of this property weakens condition $V(S)$, and consists in the tendency to induce non-monotonicity over quarrellers when the original game's vulnerability is less restrictive, namely, it contains at least one division in which, prior to the quarrel in $\mathcal{G}$, the quarrelling pair votes together and is successful and at least one of them is not decisive (whether the other is decisive or not).
We say that a conception $\mathcal{Q}$ is {\em stronlgy disposed to induce non-montonicity over quarrellers}
if, for any quarrelling pair $\{i,j\}$ and for any $\mathcal{G}$ subject to this vulnerability (condition $V'(S)$),
the transformation rule $\mathcal{Q}$ would induce non-monotonicity over at least one such division by rendering $i$ and $j$ unsuccessful in it when they quarrel in $\hat{\mathcal{G}}$ (condition $I(S)$).

\begin{quote}
\textbf{Strong Disposition to Induce Non-Monotonicity Over Quarrellers (SDNMQ):} 
The transformation rule $\mathcal{Q}$ is strongly disposed to induce non-monotonicity over quarrellers iff, 
for any pair $\{i,j\}$ where $i$ quarrels with $j$:
\begin{align*}
\forall \mathcal{G} \in \mathbb{G} : \qquad & V'(S) \ \Longrightarrow \ I(S)
\end{align*}
\qquad where:
\begin{align*}
V'(S) &:= \ \exists S \ (i,j \notin S) \ \mid \ S \ \cup \{i\} \in \mathcal{W} \ \vee \ S \cup \{j\} \notin \mathcal{W},\\
I(S) &:= \ \exists S \ (i,j \notin S) \ \mid \ (S \cup \{i\} \in \mathcal{W} \cap \hat{\mathcal{W}} \ \wedge \ S \cup \{i,j\} \notin \hat{\mathcal{W}}) \ \vee \ (S \cup \{j\} \notin \mathcal{W} \cup \hat{\mathcal{W}} \ \wedge \ S \in \hat{\mathcal{W}}).
\end{align*}
\end{quote}

\noindent We note that because $V'(S)$ is less restrictive than $V(S)$, some quarrels may satisfy DNMQ but not SDNMQ --
for example, if the quarrel induces non-monotonicity over quarrellers only when one of them is decisive.
However, SDNMQ does not strictly entail DNMQ.
This is because there may exist exotic quarrels with the following property:
although for most games they satisfy the entailment relation between antecedent and consequent for both SDNMQ and DNMQ,
nevertheless in games in which there exists a division $S_1$ that satisfies $V'(S_1)$ but not $V(S_1)$ and a different division $S_2$ that satisfies $V(S_2)$ and hence also $V'(S_2)$,
they induce non-monotonicity over quarrellers over the first division $S_1$ (consistent with SDNMQ) but not over the second (violating DNMQ).}

With this concept to hand, we can now present our first theorem,
which formalizes the intuition that a quarrel postulate based on a quarrel that induces non-monotonicity over quarrellers is unreasonable.
(Proofs for this and all subsequent theorems are presented in the Appendix.)

\begin{theorem}\label{thm:DNMQ}
If a reasonable conception of quarrelling $\mathcal{Q}$ (which satisfies CSR)
is disposed to induce non-monotonicity over quarrellers, 
then the standard quarrel postulate based on $\mathcal{Q}$ will be violated by {\bf any} measure of voting power $\Psi$.
\end{theorem}

We conclude that a quarrel postulate based on a conception of quarrelling $\mathcal{Q}$ that satisfies CSR
but also DNMQ is unreasonable,
i.e., the postulate would not be one that a reasonable measure of voting power should be expected to satisfy,
{\em because no measure could satisfy it}!
The point is {\em not} that being disposed to induce non-monotonicity over quarrellers would disqualify $\mathcal{Q}$ as a conception of what {\em quarrelling} is per se.
A transformation rule may very well satisfy the DNMQ property and still adequately articulate the core intuition of what a quarrel is.
The point is, rather, that such a conception would not be fit to serve as the basis for a reasonable {\em quarrel postulate}.%
\footnote{The same conclusion holds for SDNMQ:
\begin{theorem}\label{thm:SDNMQ}
If a reasonable conception of quarrelling $\mathcal{Q}$ (which satisfies CSR)
is strongly disposed to induce non-monotonicity over quarrellers, 
then the standard quarrel postulate based on $\mathcal{Q}$ will be violated by {\bf any} measure of voting power $\Psi$.
\end{theorem}}

Now, it is true that DNMQ is a stronger property than TNM,
since the latter is satisfied even if $\mathcal{Q}$ transforms merely a single game $\mathcal{G} \in \mathbb{G}$ to become non-monotonic,
whereas DNMQ requires that, for any quarrelling pair, $\mathcal{Q}$ induce {\em all} games $\mathcal{G} \in \mathbb{G}$ displaying the relevant vulnerability $V(S)$ to become non-monotonic.
Thus Theorem \ref{thm:DNMQ} does not strictly speaking demonstrate that {\em all} transformation non-monotonic quarrels are unsuitable for the quarrel postulate.
However, showing the unsuitability of conceptions disposed to induce non-monotonicity over quarrellers is sufficient for our purposes.
First, any quarrel that is transformation non-monotonic will be so over quarrellers for at least some quarrelling pair.
True, a transformation rule inducing a quarrel between two players could in principle result
in a game that is non-monotonic but not over {\em those two} quarrelling players.
Nevertheless, any non-monotonic game $\hat{\mathcal{G}}$ derived from an initially monotonic game $\mathcal{G}$
by a conception of a quarrel $\mathcal{Q}$ that satisfies CSR
will be non-monotonic over {\em some} pair(s) of players
who would be quarrellers relative to some binary monotonic game from which the quarrelling game could be derived via $\mathcal{Q}$:

\begin{theorem}\label{thm:non-mon}
Let $\hat{\mathcal{G}}$ be a non-monotonic game derived from a binary monotonic game $\mathcal{G'} \in \mathbb{G}$
by imposing a quarrell ${\mathcal{Q}}$ between two players $l$ and $m$, where ${\mathcal{Q}}$ satisfies CSR.
Then there exists a game $\mathcal{G} \in \mathbb{G}$ such that $\hat{\mathcal{G}}$ can be derived from ${\mathcal{G}}$
by imposing the same conception of a quarrel ${\mathcal{Q}}$ between players $i$ and $j$
and where $\hat{\mathcal{G}}$ is non-monotonic over the quarrellers $i$ and $j$.
\end{theorem}

\noindent This implies that TNM entails TNMQ for at least some pair of quarrellers.

Second, quarrels that satisfy TNMQ for some pair of quarrellers but fail to satisfy DNMQ are rather exotic,
comprising transformation rules that, rather than consisting in a set of general rules ranging over specific types of division for all games and quarrelling pairs,
instead single out specific types of game and/or quarreller for special treatment.
It is therefore not suprising that, as we shall demonstrate, all the non-monotonic conceptions we consider below satisfy DNMQ.

\subsection{Symmetry}\label{sec:symmetry2}
To be fit to serve as the basis for a reasonable (standard) quarrel postulate in general,
a conception of quarrelling must, in addition to not being disposed to induce non-monotonicity over quarrellers, be symmetric as well.
This is because asymmetric conceptions, despite not being unreasonable per se,
restrict the reasonability of a quarrel postulate based on them to a specific class of voting power measures.
On the one hand, a violation of symmetry is innocuous for evaluating {\em decisiveness measures} of a priori voting power,
such as the Penrose-Banzhaf measure, which track (full) decisiveness only.%
\footnote{For decisiveness measures (and success measures), see Laruelle and Valenciano 2005b.}
Decisiveness measures calculate a player's voting power such that only divisions in which the player is decisive count towards its voting power.
Asymmetry is innocuous for this class of a priori measures because every winning division in which a player is
{\sc yes}-decisive is mirrored by one and only one losing division in which the player is {\sc no}-decisive:
these are the two divisions in which the decisive player's vote varies but all other players' votes are held constant.
This implies that a player's a priori {\sc yes}-voting power is always equal to its a priori {\sc no}-voting power on such measures, $\psi_i^+=\psi_i^-$.
This is precisely why a player's Penrose-Banzhaf a priori voting power,
which by definition is equal to the proportion of all divisions in which the player is decisive,
can be calculated, as is typically done, on the basis of only the divisions in which the voter is {\sc yes}-{\em decisive}.

On the other hand, decisiveness measures are merely a subclass of the broader class of {\em efficacy measures} of voting power,
which calculate a player's voting power such that any division in which the player is casually efficacious may contribute to its voting power.
Asymmetry is not innocuous for those efficacy measures of voting power,
such as the Abizadeh-Vetta Recursive Measure,
that track not just decisiveness or full efficacy,
but also {\em partial} efficacy, and so base a player's voting power on divisions in which it plays {\em some} causal role in 
producing the outcome -- whether fully decisive or not.%
\footnote{On the class of efficacy measures, partial efficacy, and the Abizadeh-Vetta Recursive Measure, see Abizadeh (2022); Abizadeh and Vetta (2023).
For partial efficacy and degrees of causation, see also Braham and van Hees 2009.
More generally, for conditions that play a causal role but are not fully necessary for the effect, see McDermott 1995; Ramachandran 1997; Schaffer 2003.}
Asymmetry is not innocuous here because the partial efficacy of a successful {\sc yes}-voter in winning divisions is {\em not} mirrored in the corresponding
divisions in which the voter changes its vote but all other players' votes are held constant:
precisely because the voter is not (fully) decisive, its changed vote does not change the outcome;
the voter therefore flips from being a partially efficacious, successful {\sc yes}-voter to an inefficacious, unsuccessful {\sc no}-voter.
Therefore, for such measures, a player's a priori {\sc yes}-voting power is not necessarily equal -- as it is in decisiveness measures -- 
to its a priori {\sc no}-voting power.

The upshot is that a conception of quarrelling that violates symmetry may be fit to serve as the basis of
a reasonable quarrel postulate in the standard form for decisiveness measures,
but not for measures of voting power in general.
Alternatively, such a conception can be the basis for a reasonable quarrel postulate for efficacy measures in general,
but only for a non-standard postulate specified for either {\sc yes}- or {\sc no}-voting power only --
rather than for voting power in general as in the standard postulate.
In particular, if two voters quarrel only with respect to {\sc yes}-outcomes,
then their asymmetric {\sc yes}-quarrel could be the basis for only a reasonable {\sc yes}-voting-power quarrel postulate,
but not for a reasonable quarrel postulate in standard form.

Transformation monotonicity (or, more strictly, not being disposed to induce non-monotonicity over quarrellers) and symmetry exhaust our account of the properties a reasonable conception of quarrelling
must have to be fit to serve as the basis of a reasonable quarrel postulate.
As we shall argue in Section \ref{sec:nonrecip},
non-reciprocality does not pose any problems for the postulate's reasonability.

\section{Felsenthal and Machover's Conception}\label{sec:FM}

We are now in a position to evaluate the reasonability of the two existing structural conceptions of quarrelling in the literature,
and their fitness to serve as the basis for a reasonable quarrel postulate.
We begin here with Felsenthal and Machover's conception,
and turn to Laruelle and Valenciano's conception in Section~\ref{sec:LV}.
We shall later show how each conception fits into the general framework for a class of quarrels
we develop in Sections~\ref{sec:framework} to~\ref{sec:nonrecip}.

Felsenthal and Machover (1998: 237) base the quarrel postulate on the following conception of a quarrel.
They say that $i$ has a quarrel with $j$ if and only if
$S \in \mathcal{W}$ implies $i \notin S$ or $j \notin S$.
This is equivalent to defining the {\sc yes}-successful sets $\hat{\mathcal{W}}$ of $\hat{\mathcal{G}}$ as follows. 
For any $S \subseteq N$, set
\begin{align*}
&S \notin \hat{\mathcal{W}} 					&&\text{if } \{i,j\} \subseteq S\\
&S \in \hat{\mathcal{W}} \iff  S \in \mathcal{W} 	&&\text{otherwise}
\end{align*}

\noindent Thus any division in which both~$i$ and~$j$ vote {\sc yes} loses:
$i$ and $j$ cannot both be {\sc yes}-successful. Call this conception of a quarrel an {\em FM-quarrel} or {\em FM-rule}.

To begin, observe that an FM-quarrel satisfies YQ.
It is therefore a reasonable conception of a {\sc yes}-quarrel, since it captures the concept's core intuition.
It is also a {\em reciprocal} conception.

The FM-quarrel, however, has two shortcomings.
First, because it is restricted to a {\sc yes}-quarrel -- it does not satisfy NQ -- it violates {\em symmetry}:
an FM-quarrel models a quarrel between $i$ and $j$ over {\sc yes}-outcomes, but not over {\sc no}-outcomes.
This implies, as argued earlier, that the FM-quarrel postulate (that is, the standard quarrel postulate based on an FM-quarrel)
is not reasonable for measures of voting power in general:
even if it were reasonable to expect decisiveness measures to satisfy it, it would not be so for efficacy measures in general.

Second, as Felsenthal and Machover themselves note, the FM-quarrel violates {\em monotonicity}.
For example, suppose there exists a set $S$, containing neither $i$ nor $j$, 
such that either $S \cup \{i\} \in \mathcal{W}$ or $S \cup \{j\} \in \mathcal{W}$ or both.
Then monotonicity is violated in $\hat{\mathcal{G}}$ since $S\cup \{i,j\} \notin \hat{\mathcal{W}}$.
But perhaps this is a minor violation, and although the FM-conception sometimes produces non-monotonic games,
it is not disposed to induce non-monotonocity over quarrellers?
No, an FM-rule also satisfies DNMQ:

\begin{theorem}\label{thm:FM-DNMQ}
The FM-rule is disposed to induce non-monotonicity over quarrellers.
\end{theorem}

Two conclusions follow.
First, because it fails monotonicity and satisfies DNMQ, an FM-quarrel is not fit to be basis for the standard quarrel postulate. %
We can illustrate the fact that any measure of voting power $\Psi$ will violate the standard quarrel postulate based on the FM-rule with a simple example.
Consider dictator-rule voting with three voters, one dictator and two dummies.
Inducing an FM-quarrel between the dictator and a dummy in this voting game $\mathcal{G}$ renders the dummy decisive in half of the eight divisions in $\hat{\mathcal{G}}$.
(The dummy becomes decisive in the two divisions in which they quarrel on the {\sc yes} side,
along with the two divisions that combine with each of these to compose a violating pair.)
Second, and for precisely the same reason, some players, like the dummy in our example, may have good reasons to provoke an FM-quarrel with other players
to increase their voting power.
We explore the political significance of this in our conclusion.

Felsenthal and Machover themselves, however, draw a more sweeping conclusion from their analysis:
not merely that it would not be reasonable to expect measures of a priori voting power to satisfy a quarrel postulate based on a quarrel as they conceive it, 
but that it would not be reasonable to expect them to satisfy a quarrel postulate based on a structural conception of quarrelling {\em in general}.%
\footnote{As they put it, the paradox of quarrelling members is a merely ``superficial'' paradox (1998: 223, 241).
They also suggest that ``the concept of {\em quarrelling} oversteps the limits of the aprioristic terrain''
and so ``makes no sense'' for efficacy measures of voting power such as $PB$ (1998: 238-239).
However, as noted above, there exists a perfectly intuitive interpretation ready to hand
for structural conceptions of quarrelling (appropriate for a postulate on a priori power).}
We think this more sweeping conclusion fails to appreciate the availability of other conceptions (to be presented below) that do satisfy monotonicity.

\section{Laruelle and Valenciano's Conception}\label{sec:LV}
In contrast to Felsenthal and Machover, who abandon the quarrel postulate as unreasonable,
Laruelle and Valenciano (2005a) take the FM-quarrel's failure to satisfy monotonicity as the occasion for proposing
a revised conception of a quarrel on which to base the quarrel postulate.
On their revised account, a game $\hat{\mathcal{G}}$ in which $i$ quarrels with $j$ is 
derived from $\mathcal{G}$ by stipulating that the {\sc yes}-successful sets $\hat{\mathcal{W}}$ satisfy the properties:
\begin{align*}
S \in \hat{\mathcal{W}} &\iff S\setminus \{i\} \in \mathcal{W} &\forall S: j\in S\\
S \in \hat{\mathcal{W}} &\iff S\cup \{i\} \in \mathcal{W} &\forall S: j\notin S
\end{align*}

\noindent which we can rewrite, for any $S$ containing neither $i$ nor $j$, as:
\begin{align*}
S\cup \{i,j\} \in \hat{\mathcal{W}} &\iff S\cup \{j\}\in \mathcal{W}\\
S \in \hat{\mathcal{W}} &\iff S\cup \{i\}\in \mathcal{W}
\end{align*}

To begin, note that Laruelle and Valenciano's conception of a quarrel, which we call an {\em LV-quarrel} or {\em LV-rule}, satisfies
both CSR and {\em symmetry}.
It does, however, violate {\em reciprocality}:
the formulation models a non-reciprocal quarrel that $i$ has against $j$.
Player $i$'s LV-quarrel with player~$j$ is different from $j$'s LV-quarrel with $i$.

\begin{theorem}\label{thm:LV-nonr}
The LV-rule is a non-reciprocal quarrel.
\end{theorem}

Although the violation of reciprocality does not fit Laruelle and Valenciano's stated motivation
of modelling Brams' conception of a reciprocal quarrel,
it is not a problem for the conception per se:
as we formally demonstrate in Section~\ref{sec:nonrecip}, non-reciprocality does not disqualify a conception 
as a reasonable conception of a quarrel,
nor as a reasonable basis for a quarrel postulate.

The LV conception does, however, suffer from a critical flaw:
even though Laruelle and Valenciano propose their revision to Felsenthal and Machover's conception on the grounds 
that the latter violates monotonicity,
so too does their own proposal! 
For example, consider the voting game $\mathcal{G}$ with two players and $\mathcal{W}=\{\{1\},\{1,2\}\}$.
Now consider game $\hat{\mathcal{G}}^{1,2}$, which incorporates the quarrel of $i=1$ against $j=2$.
As shown in the proof of Theorem \ref{thm:LV-nonr}, $\hat{\mathcal{W}}^{1,2}=\{\emptyset, \{1\}\}$.
Thus $\emptyset \in \hat{\mathcal{W}}$ and $\{1\} \in \hat{\mathcal{W}}$ but $\{1,2\} \notin \hat{\mathcal{W}}$;
similar violations arise in every voting game constructed under this formulation.
Again, not only does the LV-rule violate monotonicity, it satisfies DNMQ:
\begin{theorem}\label{thm:LV-DNMQ}
The LV-rule is disposed to induce non-monotonicity over quarrellers.
\end{theorem}
\noindent An LV-quarrel therefore does not furnish the basis for a reasonable quarrel postulate.

\section{A General Quarrel Framework}\label{sec:framework}

In light of the failure of existing specifications of the (standard) quarrel postulate,
we seek a new conception of quarrelling fit to serve as the basis for a reasonable postulate.
We pursue our goal by specifying a class of quarrels allowing for a variety of {\em magnitudes} of quarrel.
This will give us a unified framework in which to compare alternate conceptions of quarrelling.
Our framework is also designed to serve our second overarching goal, namely,
to reveal the kinds of quarrel in which players can gain by quarrelling with others.

Recall that a (structural) quarrel consists in the reduction or neutralization of effective cooperation between the quarrelling players
when they vote on the same side.
Thus, given a voting game $\mathcal{G}$, to incorporate a quarrel between $i$ and $j$ we must
characterize the outcome for any $S\cup\{i,j\}$.
This is precisely what {\em cooperative-success-reduction} does:
CSR requires that, in the presence of a quarrel between players~$i$ and $j$,
any voting group containing both $i$ and $j$ not be more cooperatively successful than it would have been 
in the absence of a quarrel between players~$i$ and $j$;
and that there be at least one game with a group containing $i$ and $j$ that is less successful when they quarrel than it would have been absent the quarrel.
Thus CSR furnishes the starting point for our framework.

To sharpen our analysis, however,
we need to isolate the concept of a quarrel from other, distinct concepts,
by imposing further constraints on the transformation rule $\mathcal{Q}$.
In particular, we should distinguish imposing a quarrel from three other ways in which a voting game could be transformed.
Whereas imposing a quarrel reduces or neutralizes the effectiveness of cooperation,
inducing {\em enhanced cooperation} produces the opposite effect:
it transforms the original game by strengthening effective cooperation and therefore increasing
the cooperative success of the players whose cooperation is enhanced.
Quarrelling and enhanced cooperation both concern cooperation, or votes that {\em agree}:
the transformation operates on those divisions in which the players concerned vote on the same side.
Consider, by contrast, a transformation in which, in the derived game,
player~$i$ neutralizes the effectiveness of $j$'s vote in those divisions in which the two players' votes {\em disagree}.
This is what we call an {\em ambush} of $j$ by $i$.
Or consider the opposite of an ambush, which is induced when,
in divisions in which $i$ and $j$ vote on opposite sides, $i$ enhances the effectiveness of $j$.
We call this a {\em betrayal} (of the group with whom $i$ ostensibly cooperates) by~$i$ in favour of $j$.

A transformation of the original game formally cannot combine, for a given pair of players, a quarrel and enhanced cooperation
on the same side ({\sc yes} or {\sc no}),
but a quarrel could be combined with an ambush.
Formally, a quarrel between $i$ and $j$ could also be combined with a betrayal by $i$ in favour of $j$.
(But it is difficult to give any substantive, intuitive interpretation to such a hybrid transformation:
if $i$ is motivated to quarrel with $j$ when their votes agree,
it is difficult to see why~$i$ would be motivated to betray its own side in favour of $j$ when their votes disagree.)

To rule out hybrid transformations, which combine a quarrel between two players
with either an ambush or betrayal centred on their relationship,
we therefore also impose, in addition to CSR, the following condition on the transformation rule $\mathcal{Q}$:
\begin{align*}
\text{{\bf Non-Ambush/Betrayal ($\neg$AB).}} & \quad \text{For any $S$ containing neither $i$ nor $j$:}\\
\forall \mathcal{G}\in \mathbb{G}:\ S\cup \{i\} \in \hat{\mathcal{W}} &\iff  S\cup \{i\} \in \mathcal{W} \tag{$\neg{AB}$-1} \label{eq:nAB1}\\
\forall \mathcal{G}\in \mathbb{G}:\ S\cup \{j\} \in \hat{\mathcal{W}} &\iff  S\cup \{j\} \in \mathcal{W} \tag{$\neg{AB}$-2} \label{eq:nAB2}
\end{align*}

In \ref{eq:nAB1},
the entailment from left to right says that for any division in which $i$'s and $j$'s votes disagree and $i$ votes {\sc yes},
if the outcome is {\sc no} in $\mathcal{G}$,
then it is also {\sc no} in $\hat{\mathcal{G}}$.
This implies that $i$ is not more {\sc yes}-successful and $j$ is not less {\sc no}-successful due to the transformation when their votes disagree.
This means that $i$ does not ambush $j$ on the {\sc no} side, and (what is formally equivalent)
$j$ does not betray its own {\sc no} side in favour of $i$.
By contrast, the entailment from right to left says that for any division in which $i$'s and $j$'s votes disagree and $i$ votes {\sc yes},
if the outcome is {\sc yes} in $\mathcal{G}$,
then it is also {\sc yes} in $\hat{\mathcal{G}}$.
This implies that $i$ is not less {\sc yes}-successful and $j$ is not more {\sc no}-successful due to the transformation when their votes disagree.
This means that $j$ does not ambush $i$ on the {\sc yes} side, and (what is formally equivalent)
$i$ does not betray its own {\sc yes} side in favour of $j$.

In \ref{eq:nAB2},
the entailment from left to right says that for any division in which $i$'s and $j$'s votes disagree and $i$ votes {\sc no},
if the outcome is {\sc no} in $\mathcal{G}$,
then it is also {\sc no} in $\hat{\mathcal{G}}$.
This implies that $i$ is not less {\sc no}-successful and $j$ is not more {\sc yes}-successful due to the transformation when their votes disagree.
This means that $j$ does not ambush $i$ on the {\sc no} side, and (what is formally equivalent)
$i$ does not betray its own {\sc no} side in favour of $j$.
By contrast, the entailment from right to left says that for any division in which $i$'s and $j$'s votes disagree and $i$ votes {\sc no},
if the outcome is {\sc yes} in $\mathcal{G}$,
then it is also {\sc yes} in $\hat{\mathcal{G}}$.
This implies that $i$ is not more {\sc no}-successful and $j$ is not less {\sc yes}-successful due to the transformation when their votes disagree.
This means that $i$ does not ambush $j$ on the {\sc no} side, and (what is formally equivalent)
$j$ does not betray its own {\sc no} side in favour of $i$.

CSR and $\neg$AB together provide six formulae
that characterize our complete general framework for a pure quarrel
(that is, a quarrel not combined with an ambush or betrayal):

\begin{quote}
\text{{\bf General Quarrel Framework.}}
For any $S$ containing neither $i$ nor $j$: 
\begin{align*}
& \text{{\bf CSR: YQ}}\\
&\qquad \forall \mathcal{G}\in \mathbb{G}:\ & S\cup \{i,j\} \in \hat{\mathcal{W}} & \implies  S\cup \{i,j\} \in \mathcal{W} \tag{YQ-1} \\
&\qquad \exists \mathcal{G}\in \mathbb{G}:\ & \exists S \mid \ S\cup \{i,j\} \notin \hat{\mathcal{W}} & \ \ \wedge\ \ \ S\cup \{i,j\} \in \mathcal{W} \tag{YQ-2} \\[0.2cm]
& \text{{\bf CSR: NQ}}\\
&\qquad \forall \mathcal{G}\in \mathbb{G}:\ & S\in \hat{\mathcal{W}} &\Longleftarrow S\in \mathcal{W} \tag{NQ-1}\\
&\qquad \exists \mathcal{G}\in \mathbb{G}:\ &\exists S \mid \ S\in \hat{\mathcal{W}} &  \ \ \wedge\  S\notin \mathcal{W} \tag{NQ-2} \\
& \text{{\bf $\neg${AB}}}\\
&\qquad \forall \mathcal{G}\in \mathbb{G}:\ & S\cup \{i\} \in \hat{\mathcal{W}} & \iff  S\cup \{i\} \in \mathcal{W} \tag{$\neg${AB}-1} \\
&\qquad \forall \mathcal{G}\in \mathbb{G}:\ & S\cup \{j\} \in \hat{\mathcal{W}} & \iff  S\cup \{j\} \in \mathcal{W} \tag{$\neg${AB}-2} 
\end{align*}
\end{quote}

\noindent CSR implies that the transformation imposes a quarrel,
and $\neg$AB implies that it imposes only a quarrel.

Because CSR may be satisfied in numerous ways,
this general quarrel framework leaves a large degree of flexibility for further specification.
In particular, we can satisfy CSR by imposing more restrictive conditions that entail CSR
but represent varying degrees or magnitude of quarrel, whether reciprocal or not.
To simplify our analysis of these degrees of quarrelling,
we begin by amending our general framework,
in order to focus on quarrels on the {\sc yes} side only.
We introduce a general framework for asymmetric {\sc yes}-quarrels
by replacing NQ with the following condition:
\begin{quote}

\text{{\bf Zero {\sc No}-Quarrel.}}
For any $S$ containing neither $i$ nor $j$: 
\begin{align*}
\forall \mathcal{G}\in \mathbb{G}:\ S \in \hat{\mathcal{W}} \iff  S \in \mathcal{W} \tag{$\neg${NQ}} \label{eq:nNQ}
\end{align*}
\end{quote}

The entailment from left to right in $\neg$NQ states that for any division in which $i$ and $j$ vote {\sc no},
if the outcome is {\sc no} in $\mathcal{G}$,
then it is also {\sc no} in $\hat{\mathcal{G}}$.
This implies that $i$ and $j$ are just as {\sc no}-successful in $\hat{\mathcal{G}}$ as in $\mathcal{G}$,
which means there is no {\sc no}-quarrel between $i$ and $j$.
(This obviously negates \ref{eq:NQ2}.)
The entailment from right to left (which is obviously equal to \ref{eq:NQ1}),
states in turn that for any division in which $i$ and $j$ vote {\sc no},
if the outcome is {\sc yes} in $\mathcal{G}$,
then it is also {\sc yes} in $\hat{\mathcal{G}}$.
This implies that neither is more {\sc no}-successful in $\hat{\mathcal{G}}$ than in $\mathcal{G}$,
which means there is no {\em enhanced} cooperation on the {\sc no} side between $i$ and $j$ either.

With this simplified, asymmetric framework to hand,
we now turn to fleshing out our framework by imposing varying degrees of quarrel.
We shall do this by tightening YQ.
For the sake of clarity we begin with reciprocal quarrels.
We return to the symmetric framework in Section~\ref{sec:symm-quarrel},
and consider non-reciprocal quarrels in Section~\ref{sec:nonrecip}.
Throughout our framework, for all quarrels, we assume they are pure and hence assume $\neg$AB.

\section{Three Degrees of Quarrel}\label{sec:three-degrees}
A quarrel can adversely affect a group's ability to cooperate effectively in different ways,
depending on the {\em strength} of the quarrel.
There are three intuitive ways to characterize the strength of a quarrel: as {\em weak}, {\em strong}, and {\em cataclysmic}.

To understand these concepts note that monotonicity has a simple interpretation in voting games and in coalition games more generally:
a group or coalition $S$ is at least as effective as its most effective sub-group.
That is, if the votes of $T\subset S$ are sufficient to secure a {\sc yes}-outcome,
then the votes of $S$ are also sufficient.
Equivalently, if the sub-group $T$ is sufficient to secure a {\sc yes}-outcome,
then a group $S$ is also sufficient if the members of $T$ simply work together to the exclusion of,
or in isolation from, the members of $S\setminus T$.
The question is what impact a quarrel between two members $i$ and $j$ has on the group $S$'s ability to secure a {\sc yes}-outcome.
Assume that $i$ and $j$ have a reciprocal quarrel with each other in the group $S$. Then:
\begin{itemize}
\item A reciprocal weak quarrel renders $i$ and $j$ unable to function with each other.
\item A reciprocal strong quarrel renders $i$ and $j$ unable to function with any member of $S$.
\item A reciprocal cataclysmic quarrel renders every member of $S$ unable to function.
\end{itemize}

\noindent This gives us three degrees of quarrelling --
or four if we include no quarrelling, i.e., a quarrel of {\em zero magnitude}.
We have already formalized the notion of a zero or non-quarrel (on the {\sc no} side) with $\neg$NQ .
We now take up each of the three positive conceptions of quarrelling in turn.

\subsection{Reciprocal Weak Quarrelling}
Take any group $S\cup\{i,j\}$.
In a weak quarrel, the sub-groups $S\cup\{i\}$ and $S\cup \{j\}$ still function as well as before.
Thus the group $S\cup\{i,j\}$ may simply isolate or exclude one of $i$ or $j$ and obtain the best outcome achievable 
by just the votes of either $S\cup\{i\}$ or $S\cup \{j\}$.
We can therefore define a reciprocal weak {\sc yes}-quarrel by adopting the general asymmetric framework above,
but tightening YQ by replacing it with a further specification:
\begin{align*}
\forall \mathcal{G}\in \mathbb{G}:\ S\cup \{i,j\} \in \hat{\mathcal{W}} &\iff S\cup \{i\} \in \mathcal{W} \ \vee\  S\cup \{j\}  \in \mathcal{W} \tag{YQ\textsubscript{w}} \label{eq:YQw}
\end{align*}
\ref{eq:YQw} is a specification of YQ because, given that the original game $\mathcal{G}$ is monotonic,
satisfying \ref{eq:YQw} entails satisfying both \ref{eq:YQ1} and \ref{eq:YQ2}.
This reciprocal weak conception is therefore a reasonable conception of a {\sc yes}-quarrel:
by construction, it satisfies YQ.

The entailment from left to right states that if, for any division in which $i$ votes {\sc yes} in agreement with a set of voters excluding $j$,
the outcome is {\sc no} in $\mathcal{G}$,
{\bf and} if, for the division in which $j$ votes {\sc yes} in agreement with that same set of voters,
the outcome is also {\sc no} in $\mathcal{G}$,
then if $i$ and $j$ vote {\sc yes} in agreement with that same set of voters,
the outcome will also be {\sc no} in the derived game $\hat{\mathcal{G}}$.
This implies that $i$ and $j$ cannot be more {\sc yes}-successful in $\hat{\mathcal{G}}$ than
the most {\sc yes}-successful of~$i$ and $j$ without the other in $\mathcal{G}$.
This means that effective cooperation beween $i$ and $j$ on the {\sc yes} side is completely neutralized,
that is, $\hat{\mathcal{G}}$ comprises {\em at least} a weak {\sc yes}-quarrel.
By contrast, the entailment from right to left states that 
if, for any division in which $i$ votes {\sc yes} in agreement with a set of voters excluding $j$,
the outcome is {\sc yes} in $\mathcal{G}$,
{\bf or} if, for the division in which $j$ votes {\sc yes} in agreement with that same set of voters,
the outcome is also {\sc yes} in $\mathcal{G}$,
then if $i$ and $j$ vote {\sc yes} in agreement with that same set of voters,
the outcome will also be {\sc yes} in $\hat{\mathcal{G}}$.
This means that $\hat{\mathcal{G}}$ does not comprise a {\sc yes}-quarrel any stronger than weak:
in particular, it does not comprise a strong {\sc yes}-quarrel.

Consider what a weak quarrel so-defined implies for a weighted voting game $\mathcal{G}$,
that is, a voting game in which each player's vote has a fixed weight
and a {\sc yes}-outcome requires meeting a specified quota of total {\sc yes}-vote-weights (Felsenthal and Machover 1998: 29-32).
If $\hat{\mathcal{G}}$ contains a reciprocal weak quarrel between $i$ and $j$,
then both of them voting {\sc yes} together in $\mathbb{S}\cup\mathbbm{\{i,j\}}$
forces the weight of the {\bf weaker} of the two players to $0$. 

Importantly, reciprocal weak quarrelling satisfies transformation {\em monotonicity}:
\begin{theorem}\label{thm:a-r-w}
An asymmetric reciprocal weak {\sc yes}-quarrel is transformation monotonic.
\end{theorem}

\subsection{Reciprocal Strong Quarrelling}
In a strong quarrel, not only are $i$ and $j$ unable to cooperate effectively with each other,
$i$ and $j$ cannot individually do so with any other members of $T$ either.
It is as if their quarrel so disturbs the quarrelling pair that they are no longer in a position to cooperate effectively with anyone.
Hence the group $S\cup\{i,j\}$ can obtain the outcome achievable by just the votes of $S$.
We can therefore define a reciprocal strong {\sc yes}-quarrel by adopting the general asymmetric framework above,
but further specifying condition YQ as:
\begin{align*}
\forall \mathcal{G}\in \mathbb{G}:\ S\cup \{i,j\} \in \hat{\mathcal{W}} &\iff S\in \mathcal{W} \tag{YQ\textsubscript{s}} \label{eq:YQs}
\end{align*}
The entailment from left to right states that for any division in which $i$ and $j$ vote {\sc no},
if the outcome is {\sc no} in $\mathcal{G}$,
then for the division that is identical except both $i$ and $j$ vote {\sc yes},
the outcome in $\hat{\mathcal{G}}$ would still be {\sc no}.
This implies that if there is a group excluding $i$ and $j$ that could not secure a {\sc yes}-outcome on its own,
but could do so with the cooperation of one of $i$ or $j$,
that group could not do so in $\hat{\mathcal{G}}$ if {\em both} quarrelling players join them to vote {\sc yes}.
This means that $i$ and $j$'s {\sc yes}-quarrel is a strong quarrel.
The entailment from right to left, by contrast, states that for any division in which $i$ and $j$ vote {\sc no},
if the outcome is {\sc yes} in $\mathcal{G}$,
then for the division that is identical except both $i$ and $j$ vote {\sc yes},
the outcome in $\hat{\mathcal{G}}$ would still be {\sc yes}.
This means that if other players can secure a {\sc yes}-outcome without $i$ or $j$'s vote,
then they can do so even if $i$ and $j$ bring themselves and their quarrel to the {\sc yes} side --
by excluding or isolating the quarrelling pair.
In other words, their {\sc yes}-quarrel is not cataclysmic.

Consider what a reciprocal strong quarrel implies for a weighted voting game $\mathcal{G}$.
If $\hat{\mathcal{G}}$ contains a strong quarrel between $i$ and $j$,
then if both of them vote {\sc yes} together in $\mathbb{S}\cup \mathbbm{\{i,j\}}$,
this forces the weight of {\bf both} two players to $0$.

Again, a reciprocal strong {\sc yes}-quarrel clearly satisfies YQ:
because the original game $\mathcal{G}$ is monotonic,
satisfying \ref{eq:YQs} entails satisfying both \ref{eq:YQ1} and \ref{eq:YQ2}.
Reciprocal strong quarrelling does not, however, satisfy transformation {\em monotonicity}.
This is because a reciprocal strong quarrel affects not just the quarrellers themselves as a pair,
but also implies that neither can effectively cooperate with any other members of $S\cup \{i,j\}$ either.
Since by definition of a reciprocal strong quarrel the outcome of $\mathbb{S}\cup\mathbbm{\{i,j\}}$
is equal to the outcome of $\mathbb{S}$, monotonicity may be violated.
This can be seen with a simple example.
Consider the two-player game $\mathcal{G}$ with {\sc yes}-successful sets $\mathcal{W}=\{ \{1\}, \{2\}, \{1,2\} \}$.
When players $1$ and $2$ quarrel, we have a game $\hat{\mathcal{G}}$ where $\{1\} \in \hat{\mathcal{W}}$ and $\{2\} \in \hat{\mathcal{W}}$, but $\{1,2\} \notin \hat{\mathcal{W}}$.
Indeed, such a quarrel satisfies DNMQ:

\begin{theorem}\label{thm:a-r-s-DNMQ}
An asymmetric reciprocal strong {\sc yes}-quarrel $\mathcal{Q}$ is disposed to induce non-monotonicity over quarrellers.
\end{theorem}

The reciprocal strong {\sc yes}-quarrel therefore falls prey to to Theorem \ref{thm:DNMQ}.
This can again be illustrated by dictator-rule voting with three voters in $\mathcal{G}$.
In a quarrel between the dictator and a dummy, the dummy becomes decisive in half of the eight divisions in $\hat{\mathcal{G}}$,
thus causing any measure of voting power $\Psi$ to violate the quarrel postulate.
Such a quarrel is therefore not fit to be the basis for a reasonable (standard) quarrel postulate.

\subsection{Reciprocal Cataclysmic Quarrelling}
Intuitively, a reciprocal cataclysmic quarrel between $i$ and $j$ is a quarrel so destructive that its effects spill beyond $i$ and $j$
and prevent {\em any} members of the group from effectively cooperating to secure an outcome whenever $i$ and $j$ join their ranks.
A cataclysmic quarrel is so destructive we might dub it a {\em war}
-- a war between $i$ or $j$ that has, as it were, the collateral damage of ruining effective cooperation between
(or the efficacy of) any players who vote on the side of the quarrelling couple.
Since a reciprocal cataclysmic quarrel between $i$ and $j$ renders the group $S\cup\{i,j\}$
so ineffective that no members can function properly;
the group must exclude all its members to function!
Hence the group $S\cup\{i,j\}$ can obtain the outcome achievable by just the votes of $\emptyset$.
We can therefore define a cataclysmic {\sc yes}-quarrel by adopting the general asymmetric framework,
but further specifying condition YQ as:
\begin{align*}
\forall \mathcal{G}\in \mathbb{G}:\ S\cup \{i,j\} \in \hat{\mathcal{W}} &\iff \emptyset \in \mathcal{W} \tag{YQ\textsubscript{c}} \label{eq:YQc}
\end{align*}
This implies that a reciprocal cataclysmic {\sc yes}-quarrel is precisely Felsenthal and Machover's conception of a quarrel!

The entailment from left to right states that
for the division in which everyone votes {\sc no},
if the outcome is {\sc no} in $\mathcal{G}$,
then for any division in which $i$ and $j$ vote {\sc yes},
the outcome is {\sc no} in $\hat{\mathcal{G}}$ --
even if everyone else were to vote {\sc yes}.
This means that if $i$ and $j$ vote {\sc yes}, their quarrel deluges any other {\sc yes}-voters and renders them ineffective,
i.e., this is a cataclysmic {\sc yes}-quarrel.
The entailment from right to left states that
for the division in which everyone votes {\sc no},
if the outcome is {\sc yes} in $\mathcal{G}$ (which of course never holds if $\mathcal{G}$ is non-trivial in addition to monotonic),
then for any division in which $i$ and $j$ vote {\sc yes},
the outcome is {\sc yes} in $\hat{\mathcal{G}}$.
This implies that $i$ and $j$ are not less {\sc yes}-successful in $\hat{\mathcal{G}}$ than $\emptyset$,
which means that if even $\emptyset$ is sufficient to secure a {\sc yes}-outcome,
then even a war between $i$ and $j$ cannot prevent {\sc yes}.

The nature of this definition is illuminated by considering its effect on a weighted voting game $\mathcal{G}$.
If $\hat{\mathcal{G}}$ contains a reciprocal cataclysmic quarrel between $i$ and $j$,
then when they both vote {\sc yes} together in $\mathbb{S}\cup \mathbbm{\{i,j\}}$,
this forces the weight of {\bf every} player in $S\cup\{i,j\}$ to $0$.

Although cataclysmic {\sc yes}-quarrelling is a rather exaggerated type of quarrel,
it is a perfectly reasonable conception:
because satisfying \ref{eq:YQc} entails satisfying \ref{eq:YQ1} and \ref{eq:YQ2},
a reciprocal cataclysmic {\sc yes}-quarrel satisfies YQ.
But, as we have already proven in Theorem~\ref{thm:FM-DNMQ} when discussing the FM-conception,
an asymmetric reciprocal cataclysmic quarrel not only violates monotonicity, it satisfies DNMQ.

\begin{corollary}\label{cor:c-a-r}
An asymmetric reciprocal cataclysmic {\sc yes}-quarrel is disposed to induce non-monotonicity over quarrellers. \qed
\end{corollary}

\subsection{Symmetric Quarrelling}\label{sec:symm-quarrel}
The degrees of quarrel we have analyzed for the {\sc yes} side apply equally to the {\sc no} side.
Therefore, with our classification by degree of quarrel in hand,
we can now return to our General Quarrel Framework,
and specify it further, by degree of quarrel, to yield three positive degrees of symmetric reciprocal quarrel.
To do so, for each type of quarrel we must replace YQ and NQ with the corresponding
weak, strong, or cataclysmic specification.
Each type of symmetric reciprocal quarrel is therefore equal to the General Quarrel Framework,
except we replace YQ and NQ in each case, respectively, with the following.
Note again that throughout our framework we are characterizing pure quarrels and so assume that condition $\neg$AB holds, i.e., in each instance we assume that the quarrel does not affect divisions in which the quarrellers do not vote together.
For all $\mathcal{G} \in \mathbb{G}$ and for any $S$ containing neither $i$ nor $j$:
\begin{align*}
\text{{\bf Symmetric Reciprocal Weak Quarrel:}}\\
S\cup \{i,j\} \in \hat{\mathcal{W}} &\iff S\cup \{i\} \in \mathcal{W} \ \vee\  S\cup \{j\}  \in \mathcal{W} \tag{YQ\textsubscript{w}} \\
S \in \hat{\mathcal{W}} &\iff S\cup \{i\} \in \mathcal{W} \ \wedge\  S\cup \{j\}  \in \mathcal{W} \tag{NQ\textsubscript{w}} \label{eq:NQw}\\
\text{{\bf Symmetric Reciprocal Strong Quarrel:}}\\
S\cup \{i,j\} \in \hat{\mathcal{W}} &\iff S\in \mathcal{W} \tag{YQ\textsubscript{s}} \\
S \in \hat{\mathcal{W}} &\iff S\cup \{i,j\}\in \mathcal{W} \tag{NQ\textsubscript{s}} \label{eq:NQs}\\
\text{{\bf Symmetric Reciprocal Cataclysmic Quarrel:}}\\
S\cup \{i,j\} \in \hat{\mathcal{W}} &\iff \emptyset \in \mathcal{W} \tag{YQ\textsubscript{c}} \\
S \in \hat{\mathcal{W}} &\iff N \in \mathcal{W} \tag{NQ\textsubscript{c}} \label{eq:NQc}
\end{align*}

We have already furnished the verbal statement and intuitive meaning for the {\sc yes} side.
For the {\sc no} side, in \ref{eq:NQw}, the entailment from left to right means that $\hat{\mathcal{G}}$ does not comprise a {\sc no}-quarrel any stronger than weak:
in particular, it does not comprise a strong {\sc no}-quarrel.
The entailment from left to right, in turn, means that effective cooperation
between $i$ and $j$ on the {\sc yes} side is completely neutralized,
that is, that $\hat{\mathcal{G}}$ comprises {\em at least} a weak {\sc no}-quarrel.
In \ref{eq:NQs}, the entailment from left to right means that if other players can secure a {\sc no}-outcome without $i$ or $j$'s vote,
then they can do so even if $i$ and $j$ bring themselves and their quarrel to the {\sc no} side,
that is, their {\sc no}-quarrel is not cataclysmic.
The entailment from left to right, in turn, means that $\hat{\mathcal{G}}$ does comprise a strong {\sc no}-quarrel.
Finally, in \ref{eq:NQc}, the entailment from left to right means that if even $N$ cannot prevent a {\sc no}-outcome,
then even a war between $i$ and $j$ on the {\sc no} side will not stop a {\sc no} outcome.
The entailment from right to left, in turn, means that if $i$ and $j$ bring their quarrel to the {\sc no} side,
it deluges any other {\sc no}-voters and renders them ineffective,
i.e., this is a cataclysmic {\sc no}-quarrel.

Now, just as in the asymmetric case, a symmetric reciprocal weak quarrel satisfies monotonicity:
\begin{theorem}\label{thm:s-r-w}
A symmetric reciprocal weak quarrel is transformation monotonic.
\end{theorem}

However, if a symmetric reciprocal quarrel is either strong or cataclysmic, it not only violates monotonicity, but also satisfies DNMQ:

\begin{theorem}\label{thm:s-r-s}
A symmetric reciprocal strong quarrel is disposed to induce non-monotonicity over quarrellers.
\end{theorem}
\begin{theorem}\label{thm:s-r-c}
A symmetric reciprocal cataclysmic quarrel is disposed to induce non-monotonicity over quarrellers.
\end{theorem}

\section{Non-Reciprocal Quarrelling}\label{sec:nonrecip}

Although we have treated only reciprocal quarrels up to now, our General Quarrel Framework encompasses,
and can be easily used to formalize, non-reciprocal quarrels as well.
This is important for at least two reasons.
First, it demonstrates that our framework encompasses LV-quarrels, which, as we have previously shown, are non-reciprocal.
Second, it enables us to investigate whether it could ever be advantageous for a player to engage {\em unilaterally} in a quarrel that remains unreciprocated.
It turns out that it is.
We explore this possibility in our conclusion.

We shall focus on a quarrel that $i$ has with $j$ but not vice versa.
Again, think of $j$ as entirely indisposed to quarrelling:
no matter how much others around $j$ quarrel, with $j$ itself or with others, $j$ continues to be 
willing and able to cooperate effectively with whoever effectively reciprocates cooperation with it.
But of course if $i$ quarrels with $j$, even non-reciprocally,
then $i$ does not effectively cooperate with $j$.

We here provide the specification of YQ and NQ for a non-reciprocal weak, strong, and cataclysmic quarrel, respectively:
\begin{align*}
\text{{\bf Non-Reciprocal Weak Quarrel:}}\\
S\cup \{i,j\} \in \hat{\mathcal{W}} &\iff S\cup \{j\} \in \mathcal{W} \ \vee\  S\cup \{i\}  \in \mathcal{W}
\tag{$\protect \overrightarrow{\text{YQ}}\textsubscript{w}$} \label{eq:YQiw}\\
S \in \hat{\mathcal{W}} &\iff S\cup \{j\} \in \mathcal{W} \ \wedge\  S\cup \{i\}  \in \mathcal{W}
\tag{$\protect \overrightarrow{\text{NQ}}\textsubscript{w}$} \label{eq:NQiw}\\
\text{{\bf Non-Reciprocal Strong Quarrel:}}\\
S\cup \{i,j\} \in \hat{\mathcal{W}} &\iff S\cup \{j\}\in \mathcal{W}
\tag{$\protect \overrightarrow{\text{YQ}}\textsubscript{s}$} \label{eq:YQis}\\
S \in \hat{\mathcal{W}} &\iff S\cup \{i\}\in \mathcal{W}
\tag{$\protect \overrightarrow{\text{NQ}}\textsubscript{s}$} \label{eq:NQis}\\
\text{{\bf Non-Reciprocal Cataclysmic Quarrel:}}\\
S\cup \{i,j\} \in \hat{\mathcal{W}} &\iff \{j\} \in \mathcal{W}
\tag{$\protect \overrightarrow{\text{YQ}}\textsubscript{c}$} \label{eq:YQic}\\
S \in \hat{\mathcal{W}} &\iff N\setminus\{j\} \in \mathcal{W}
\tag{$\protect \overrightarrow{\text{NQ}}\textsubscript{c}$} \label{eq:NQic}
\end{align*}
 
The first thing to note is that a non-reciprocal weak quarrel is formally identical to a reciprocal weak quarrel.
It immediately follows that it is monotonic, by Theorem~\ref{thm:a-r-w} and Theorem~\ref{thm:s-r-w}.
\begin{corollary}\label{cor:sa-n-w}
A (symmetric or asymmetric) non-reciprocal weak quarrel is transformation monotonic. \qed
\end{corollary}
\noindent The only thing that changes is the quarrel's substantive meaning, i.e., the interpretation we give to the formal model:
whereas in a reciprocal quarrel neither $i$ nor $j$ is willing or able to cooperate effectively with the other,
in a non-reciprocal quarrel of $i$ against $j$, only $i$ is not willing or able to cooperate effectively with $j$;
$j$ is in principle willing and able, but does not effectively cooperate with $i$ because effective cooperation cannot be unilateral.

When we turn to a strong quarrel, by contrast, non-reciprocality does make a formal difference.
This is because a strong quarrel reduces the cooperative efficacy of the actively quarrelling players with {\em everyone}
else whose vote agrees with the quarrelling player's vote.
But in a non-reciprocal quarrel,
the player $j$ is not actively quarrelling, and so, unlike in a reciprocal strong quarrel,
can continue to cooperate effectively with the others despite $i$'s presence -- just as the other players can.
This is what \ref{eq:YQis} and \ref{eq:NQis} express.

The second thing to note is that this symmetric non-reciprocal strong quarrel is precisely the Laruelle and Valenciano conception of a quarrel!
It is therefore non-monotonic; indeed, by Theorem~\ref{thm:LV-DNMQ}, it satisfies DNMAQ.
\begin{corollary}\label{cor:s-n-s}
A symmetric non-reciprocal strong quarrel is disposed to induce non-monotonicity over quarrellers. \qed
\end{corollary}

Laruelle and Valenciano (2005a: 30-31) claim to have proven that no decisiveness measure of voting power
displays the quarrelling paradox based on this conception,
which, if true, would seem to call into question our claim that any (standard) quarrel postulate based on a conception of quarrelling disposed to induce non-monotonicity over quarrellers is unreasonable.
The problem with their proof, however, is that it relies on a formalization of the notion of decisiveness that is suited only for monotonic games,
i.e., it counts a voter as decisive only if its vote agrees with the outcome (2005a: 22).
They do this presumably because they did not realize that the LV-quarrel, like the FM-quarrel, also violates monotonicity.
But as noted earlier, in non-monotonic games, players may have an incentive to vote strategically (i.e., vote against the outcome they prefer)
and can be decisive even in divisions in which they have voted against the outcome.
And once we adopt the formalization of decisiveness broad enough to cover non-monotonic games,
their proof fails, as we can see with two simple counterexamples.
Consider the three-player monotonic binary voting game for which $\mathcal{W}=\{\{1,2,3\}, \{1,2\}, \{1,3\}\}$.
The Penrose-Banzhaf measure yields an a priori voting power for player 2 equal to $\frac{1}{4}$,
since it is decisive in two of the eight logically possible divisions.
If we impose an LV-quarrel of player $1$ against player $2$, however,
then player $2$'s Penrose-Banzhaf measure rises to $\frac{1}{2}$, in violation of the quarrel postulate
based on the LV-quarrel.
Even more starkly, in dictator-rule voting with three players,
if the dictator engages in an LV-quarrel against one of the dummies,
then the dummy is transformed into an (anti-)dictator.

Consider now an asymmetric non-recripocal strong {\sc yes}-quarrel of $i$ against $j$.
Again, it is easy to verify that such a quarrel does not satisfy monotonicity.
Indeed, like the reciprocal version, it also satisfies DNMQ:

\begin{theorem}\label{thm:a-n-s-DNMQ}
An asymmetric non-reciprocal strong {\sc yes}-quarrel $\mathcal{Q}$ is disposed to induce non-monotonicity over quarrellers.
\end{theorem}
\noindent Such a quarrel is therefore not fit to be basis for a reasonable (standard) quarrel postulate.

Finally, consider non-reciprocal cataclysmic quarrels.
Again here there is a difference between a non-reciprocal cataclysmic quarrel and a reciprocal one. 
Unlike all the other players whose votes agree with $i$'s,
the non-quarrelling player $j$ is unaffected by $i$'s cataclysmic quarrel with it:
if $j$ is a dictator in the original game $\mathcal{G}$,
i.e., $\{j\} \in \mathcal{W}$ and $N \setminus \{j\} \notin \mathcal{W}$,
then it can continue to dictate the outcome no matter how cataclysmically $i$ wages war in $j$'s presence.
This means the non-quarrelling player $j$ not only does not quarrel, but is immune to any quarrel --
even a quarrel directed against itself.

The non-reciprocal version, like the reciprocal one, violates non-monotonicity and satisfies DNMQ, for both the asymmetric and symmetric cases.

\begin{theorem}\label{thm:a-n-c-DNMQ}
An asymmetric non-reciprocal cataclysmic {\sc yes}-quarrel $\mathcal{Q}$ is disposed to induce non-monotonicity over quarrellers.
\end{theorem}

\begin{theorem}\label{thm:s-n-c-DNMQ}
A symmetric non-reciprocal cataclysmic quarrel $\mathcal{Q}$ is disposed to induce non-monotonicity over quarrellers.
\end{theorem}

It follows that a non-reciprocal cataclysmic quarrel, whether symmetric or asymmetric, is not fit to serve as the basis for a reasonable quarrel postulate.

\section{Conclusion: A Reasonable Quarrel Postulate and Advantageous Quarrelling}\label{sec:reasonable-QP}\label{sec:conclusion}

We have argued that the standard (structural) quarrel postulate, to be reasonable and fitting for measures of a priori voting power in general,
must be based on a conception of a quarrel that is not only itself reasonable, but also monotonic and symmetric -- unlike previous proposals.
We have accordingly developed a general framework of twelve conceptions of quarrelling,
where each conception is distinguished according to its symmetry, reciprocality, and strength in quarrelling (summarized in Table \ref{table:1}).
This framework not only encompasses the previous proposals,
it has also enabled us to identify a new conception of quarrelling (shaded in Table \ref{table:1}) that is both monotonic and symmetric,
namely, a symmetric weak quarrel, whether reciprocal or non-reciprocal (which two conceptions are, as we have shown, formally identical).
A symmetric weak quarrel is the only conception in our framework that satisfies monotonicity and concerns both {\sc yes}- and {\sc no}-voting power.
It therefore furnishes the basis for a reasonable structural quarrel postulate in its standard form --
indeed, it is the only conception in our framework to do so.

\begin{table}[h!]
\centering
\begin{tabular}{|c|| c| c| c|c|} 
 \hline
{\bf Quarrel} 
& \makecell{{\bf Asymmetric}\\  {\bf Non-Reciprocal}}
& \makecell{{\bf Asymmetric}\\  {\bf Reciprocal}}
& \makecell{{\bf Symmetric}\\  {\bf Non-Reciprocal}}
& \makecell{{\bf Symmetric}\\  {\bf Reciprocal}}\\ 
 \hline\hline
{\bf Weak} 
&  \makecell{{\small Monotonic}\\{\scriptsize [Thm~\ref{thm:a-r-w}+Cor~\ref{cor:sa-n-w}]}}  
&  \makecell{ {\small Monotonic}\\{\scriptsize [Thm~\ref{thm:a-r-w}]}}
& \cellcolor{red!15}\colorbox{red!15}{\makecell{{\small Monotonic}\\ {\scriptsize [Thm~\ref{thm:s-r-w}+Cor~\ref{cor:sa-n-w}]}}}
& \cellcolor{red!15}\colorbox{red!15}{\makecell{{\small Monotonic}\\ {\scriptsize [Thm~\ref{thm:s-r-w}]}}}\\ 
\hline
{\bf Strong} 
& \makecell{{\small DNMQ}\\{\scriptsize [Thm~\ref{thm:a-n-s-DNMQ}]}} 
& \makecell{{\small DNMQ}\\{\scriptsize [Thm~\ref{thm:a-r-s-DNMQ}]}} 
& \makecell{{\small DNMQ}\\ {\scriptsize [Thm~\ref{thm:LV-DNMQ}+Cor~\ref{cor:s-n-s}]}\\ {\scriptsize [LV-quarrel]}}
& \makecell{{\small DNMQ}\\ {\scriptsize [Thm~\ref{thm:s-r-s}]}}\\
\hline
{\bf Cataclysmic} 
& \makecell{{\small DNMQ}\\{\scriptsize [Thm~\ref{thm:a-n-c-DNMQ}]}}
& \makecell{{\small DNMQ}\\{\scriptsize [Thm~\ref{thm:FM-DNMQ}+Cor~\ref{cor:c-a-r}]} \\{\scriptsize [FM-quarrel]}}
& \makecell{{\small DNMQ}\\{\scriptsize [Thm~\ref{thm:s-n-c-DNMQ}]}} 
& \makecell{{\small DNMQ}\\{\scriptsize [Thm~\ref{thm:s-r-c}]}}\\
 \hline
\end{tabular}
\caption{Typology of a Class of Quarrels}
\label{table:1}
\end{table}

Although we have not treated hybrid quarrels here,
the power of our framework is reflected in the fact that it permits us to classify, formally characterize, and investigate these as well.
First, our framework would allow us to formalize and investigate a class of quasi-symmetric quarrels
in which voters $i$ and $j$ quarrel with each other on both the {\sc yes} and {\sc no} sides,
but in which their {\sc yes}-quarrel is different in degree from their {\sc no}-quarrel.
Similarly, it would allow us to formalize and investigate a class of quasi-reciprocal quarrels
in which both $i$ and $j$ quarrel with each other,
but in which $i$'s quarrel against $j$ is different in degree from $j$'s quarrel against $i$.
Finally, it would allow us to formalize and investigate a class of quasi-symmetric, quasi-reciprocal quarrels
in which the degree of quarrel differs not only on the {\sc yes} and {\sc no} sides,
but also between the two players.

We are now in a position to assess how the classic measures of voting power fare against the quarrel postulate based on a symmetric, weak quarrel.

\begin{theorem}\label{thm:SS}
The Shapley-Shubik index satisfies the standard 
quarrel postulate based on a symmetric, weak quarrel.
\end{theorem}

\begin{theorem}\label{thm:PB}
The Penrose-Banzhaf measure satisfies the standard 
quarrel postulate based on a symmetric, weak quarrel.
\end{theorem}

\noindent These findings are novel because neither the Shapley-Shubik index nor the Penrose-Banzhaf measure
satisfies the quarrel postulate based on the non-monotonic conceptions of quarrel used by previous scholars --
since, as we have shown, {\em any} measure of voting power will violate such a postulate.
As we have argued, a postulate that any possible measure would violate does not furnish
a reasonable standard by which to judge proposed measures.
We have shown, by contrast, that the classic measures withstand the test of the quarrel postulate reasonably construed.
Of course we have also argued that even if the Shapely-Shubik index had violated the quarrel postulate, this would not count against it,
because it is a relative, not absolute, measure.
But the Penrose-Banzhaf measure is an absolute measure, and our finding that it satisfies the quarrel postulate is both novel and significant.

Finally, our framework does not merely enable us to specify a reasonable quarrel postulate and hence a normative standard by which to assess measures of voting power.
It also pinpoints the types of quarrel that would enable a player to increase its voting power.
In particular, strong and cataclysmic quarrels can do this for the very same reason that they are an unsuitable basis for the quarrel postulate:
they are disposed to induce non-monotonicity over quarrellers.
They are thus of tremendous political significance:
they reveal incentives political actors may have to pursue conflict strategies.
Our framework predicts that actors sometimes have strategic reasons to provoke conflicts with their adversaries under the guise of cooperation
in order to undermine the effectiveness of {\em other} actors' cooperation with them or with each other.
In other words, it furnishes a formal model of the effectiveness of strategic internal {\em sabotage},
which in our model takes the form of voting in favour of outcomes one disprefers in order to undermine the effectiveness of those with whom one is ostensibly cooperating by inciting conflict with them.

Begin with strong quarrels.
Recall that when these break out, the quarrelling players cannot effectively cooperate with any other player.
Now consider the three-player simple voting game in which the first player is a dummy
and the outcome is {\sc yes} if and only if the other two both vote {\sc yes}, $\mathcal{W}=\{\{1,2,3\}, \{2,3\}\}$.
If the dummy player $1$ wages a non-reciprocal strong quarrel (whether symmetric or not) against $2$, nothing changes,
and so $1$ remains a dummy.
But if it incites player~$2$ to strongly quarrel with it -- whether $1$ reciprocates or not --
then this immediately increases $1$'s voting power: it is no longer a dummy.%
\footnote{In particular, if $2$ strongly quarrels with $1$ without reciprocation -- whether asymmetrically, let us say on the {\sc yes} side, or symmetrically --
or if they both undertake a strong asymmetric reciprocal quarrel on the {\sc yes} side,
then the quarrel transforms $\{1,2,3\}$ into a losing coalition, leaving $\hat{\mathcal{W}}=\{2,3\}$.
Here $1$ becomes decisive in both of the associated divisions.
And if $1$ and $2$ become embroiled in a strong reciprocal symmetric quarrel, then $\hat{\mathcal{W}}=\{\{2,3\}, \{3\}\}$,
in which case $1$ becomes decisive in four divisions (those associated with these two and with $N$ and $\{\emptyset\}$).}
Thus player 1 has tremendous incentives to goad 2 into strongly quarrelling with it.
This means that sometimes provoking another's wrath may be advantageous if the quarrel's effect is to render one's adversary incapable of effectively contributing to outcomes in one's presence.
In particular, the player has an incentive to vote insincerely, for an outcome its disprefers,
all the while sabotaging the outcome for which it votes by provoking a quarrel.

Cataclysmic quarrels furnish an even more dramatic case, because here a player {\em can gain even by quarrelling against another on its own, non-reciprocally}.
Consider again the three-player simple voting game with $\mathcal{W}=\{\{1,2,3\}, \{2,3\}\}$ where player $1$ is a dummy.
{\em Any} cataclysmic quarrel involving player $1$ and another player will transform $1$ into a non-dummy.%
\footnote{An asymmetric cataclysmic quarrel, let us say on the {\sc yes} side, of $2$ against $1$, whether non-reciprocally or recprocated by $1$,
transforms $\{1,2,3\}$ into a losing coalition, and so transforms player $1$ into a non-dummy decisive in two divisions.
If the player $2$ undertakes a non-reciprocal symmetric cataclysmic quarrel against $1$, then $\hat{\mathcal{W}}=\{\{2,3\}, \{3\}\}$, transforming $1$ from a dummy into a player decisive in four divisions.
And if $1$ reciprocates, then $\hat{\mathcal{W}}=\{\{2,3\}, \{3\}, \{\emptyset\}\}$, making the previous dummy decisive in six divisions!}
Most strikingly, even if $1$ unilaterally underakes a non-reciprocal cataclysmic quarrel against $2$ -- whether asymmetric or symmetric --
it becomes decisive in two divisions (having transformed the set of winning coalitions into $\hat{\mathcal{W}}=\{2,3\}$ only).
This means that $1$ can gain by acting unilaterally: {\em even if $2$ does not reciprocate}, $1$ gains in voting power by quarrelling with it.
How can this happen?
Because in a non-reciprocal cataclysmic quarrel of $i$ against $j$, even though $j$ is willing and able to cooperate effectively with others,
the quarrel so disturbs the others that none can effectively cooperate with $j$ in $i$'s presence.
Player $i$ is effectively a saboteur in the voting coalition, a player who ostensibly votes with the coalition but, by attacking a coalition member $j$,
completely undermines the coalition's capacity to function.
Politically, this is precisely what saboteurs might do to stronger adversaries: join them but sow conflict amongst their comrades.




\newpage

\section*{Appendix of Proofs}

\subsection{Proofs for Section~\ref{sec:postulate}}

\restatetheorem{thm:DNMQ}
   If a reasonable conception of quarrelling $\mathcal{Q}$ (which satisfies CSR) is disposed to induce non-monotonicity over quarrellers, 
then the standard quarrel postulate based on $\mathcal{Q}$ will be violated by {\bf any} measure of voting power $\Psi$.
\end{theorem}

\begin{proof}
For any measure $\Psi$ (which, by definition, satisfies the dummy postulates),
$\psi_j$=0 if and only if $j$ is a dummy.
We need to furnish an instance in which a dummy $j$ in the initial voting game $\mathcal{G}$
is transformed into a non-dummy in the game derived by imposing a quarrel between $i$ and~$j$
via a transformation rule $\mathcal{Q}$ that satisfies DNMQ.

Let $\mathcal{G}$ be a monotonic binary voting game in which player $i$ is a dictator and player $j$ a dummy.
Thus $\mathcal{G}$ satisfies condition $V(S)$ in DNMQ.
Now induce a quarrel between $i$ and $j$ that satisfies CSR, but which is disposed to induce non-monotonicity over quarrellers.
Then there must be at least one pair of divisions in the derived game $\hat{\mathcal{G}}$ such that:
the quarrelling players' votes agree in the first division and disagree in the second,
where the second is identical to the first except for player $j$'s vote, and (by condition $V(S)$)
in the division in which they disagree, $i$ is successful and $j$ unsuccessful;
and (by condition $I(S)$) in the division in which players $i$ and $j$ agree, they are both unsuccessful.
But since the two divisions are identical except for $j$'s vote,
it follows that $j$ is decisive in these two divisions in $\hat{\mathcal{G}}$.
Therefore, $\psi_j=0$ but $\hat{\psi}_j > 0$, in violation of the standard quarrel postulate.%
\footnote{The proof evidently relies on the fact that in non-monotonic games a player may be decisive
even when its vote disagrees with the outcome.}
\end{proof}

\restatetheorem{thm:SDNMQ}
   If a reasonable conception of quarrelling $\mathcal{Q}$ (which satisfies CSR) is strongly disposed to induce non-monotonicity over quarrellers, 
then the standard quarrel postulate based on $\mathcal{Q}$ will be violated by {\bf any} measure of voting power $\Psi$.
\end{theorem}

\begin{proof}
The proof is identical to that of Theorem \ref{thm:DNMQ}, replacing DNMQ and $V(S)$ with SDNMQ and $V'(S)$.
\end{proof}

\restatetheorem{thm:non-mon}
Let $\hat{\mathcal{G}}$ be a non-monotonic game derived from a binary monotonic game $\mathcal{G'} \in \mathbb{G}$
by imposing a quarrell ${\mathcal{Q}}$ between two players $l$ and $m$, where ${\mathcal{Q}}$ satisfies CSR.
Then there exists a game $\mathcal{G} \in \mathbb{G}$ such that $\hat{\mathcal{G}}$ can be derived from ${\mathcal{G}}$
by imposing the same conception of a quarrel ${\mathcal{Q}}$ between players $i$ and $j$
and where $\hat{\mathcal{G}}$ is non-monotonic over the quarrellers $i$ and $j$.
\end{theorem}

\begin{proof}
So $\hat{\mathcal{G}}$ is non-monotonic.
We have two cases. First, suppose the maximum cardinality subset $S$ that exhibits non-monotonicity in $\hat{\mathcal{G}}$
has $|S|=1$, i.e., $S=\{i\}$, such that $\emptyset \in \hat{\mathcal{W}}$ but $\{i\} \notin \hat{\mathcal{W}}$.
Let $\mathcal{G}$ agree with $\hat{\mathcal{G}}$ on every division except $(\emptyset, N)$.
In particular, $\emptyset \notin \mathcal{W}$. Thus $\mathcal{G}$ is monotonic.
Now take any $j\neq i$. Observe the $i$ and $j$ are both on the {\sc no}-side of $(\emptyset, N)$.
It follows that the derivation of $\hat{\mathcal{G}}$ from $\mathcal{G}$ by imposing a quarrel between $i$ and $j$ does satisfy $CSR$.
Furthermore, $\hat{\mathcal{G}}$ is non-monotonic over the quarrelling pair $\{i,j\}$, as desired.

Otherwise, there exists a pair $i,j\in N$ and a subset $S\subseteq N$ such that $S \cup \{i\} \in \hat{\mathcal{W}}$
but $S\cup\{i,j\} \notin \hat{\mathcal{W}}$.
We construct a monotonic $\mathcal{G}$ as follows.
Take any set $T\subseteq N$. \\
(1) Suppose $T\cap \{i,j\}\neq \emptyset$. If there exists an $X\subseteq T$ such that $X \in \hat{\mathcal{W}}$, then let $T \in \mathcal{W}$.\\
(2) Suppose $T\cap \{i,j\}= \emptyset$. If there exists an $X\supseteq T$ such that $X \notin \hat{\mathcal{W}}$,
then let $T \notin \mathcal{W}$.\\
(3) Otherwise let $X \in \hat{\mathcal{W}}$ if and only if $X \in \mathcal{W}$.\\
Observe that the derivation of $\hat{\mathcal{G}}$ from $\mathcal{G}$ satisfies CSR with respect to the
quarrelling pair $\{i,j\}$. Furthermore, by definition of $S$, $\hat{\mathcal{G}}$ is non-monotonic over the 
quarrelling pair $\{i,j\}$.
It remains to show that $\mathcal{G}$ is monotonic.
Suppose not. Then there is a set $T$ such that $T \in \mathcal{W}$ and $T \cup \{k\} \notin \mathcal{W}$, for some $k\notin T$.
We have three cases:

(i) Suppose $T\cap \{i,j\}\neq \emptyset$.
Since $T \cup \{k\} \notin \mathcal{W}$,
by Rule (1) there is no $X\subseteq T\cup \{k\}$ such that $X \in \hat{\mathcal{W}}$.
In particular, $T \notin \hat{\mathcal{W}}$.
But $T \in \mathcal{W}$,
so by Rule (1), there exists an $X\subseteq T$ such that $X \in \hat{\mathcal{W}}$, a contradiction.

(ii) Suppose $T\cap \{i,j\}= \emptyset$ and $k\in \{i,j\}$. Without loss of generality $k=i$. 
Then $(T\cup\{k\})\cap \{i,j\}\neq \emptyset$.
Since $T \cup \{i\} \notin \mathcal{W}$,
by Rule (1) there is no $X\subseteq T\cup \{i\}$ such that $X \in \hat{\mathcal{W}}$.
In particular, $T \notin \hat{\mathcal{W}}$.
But $T \in \mathcal{W}$,
so by Rule (2) there does not exist an $Y\supseteq T$ such that $Y \notin \hat{\mathcal{W}}$.
In particular, $T \cup \{i\} \in \hat{\mathcal{W}}$.
But then by Rule (1), $T \cup \{i\} \in \mathcal{W}$, a contradiction.

(iii) Suppose $(T\cup\{k\})\cap \{i,j\}= \emptyset$. In particular $k\notin \{i,j\}$.
Since $T \in \mathcal{W}$,
by Rule (2) there does not exist a $Y\supseteq T$ such that $Y \notin \hat{\mathcal{W}}$.
In particular, $T \cup \{k\} \in \hat{\mathcal{W}}$.
But $T \cup \{k\} \notin \mathcal{W}$,
which implies $(T\cup\{k\})\cap \{i,j\}\neq \emptyset$, a contradiction.
\end{proof}

\subsection{Proofs for Section~\ref{sec:FM}}

\restatetheorem{thm:FM-DNMQ}
The FM-rule is disposed to induce non-monotonicity over quarrellers.
\end{theorem}
\begin{proof}
Take any pair of quarrellers $\{i,j\}$ and any game $\mathcal{G}\in \mathbb{G}$ that satisfies condition $V(S)$ of DNMQ.
There thus exists an $S \ (i,j \notin S)$ where $S\cup \{i\} \in \mathcal{W}$.
Now consider an FM-quarrel $\mathcal{Q}$.
Since $\emptyset \notin \mathcal{W}$, by the FM-rule we have $S \cup \{i,j\} \notin \hat{\mathcal{W}}$.
Therefore, we have $S \cup \{i\} \in \hat{\mathcal{W}}$ and $S \cup \{i,j\} \notin \hat{\mathcal{W}}$ and so
$\mathcal{Q}$ is disposed to induce non-monotonicity over quarrellers.
\end{proof}

\subsection{Proofs for Section~\ref{sec:LV}}

\restatetheorem{thm:LV-nonr}
The LV-rule is a non-reciprocal quarrel.
\end{theorem}
\begin{proof}
Take the two-player simple voting game $\mathcal{G}$ for which $\mathcal{W} = \{\{1\}, \{1,2\}\}$.
Now suppose (contrary to fact) that the LV-rule derives a reciprocal quarrel $\hat{\mathcal{G}}$ from $\mathcal{G}$,
such that $\hat{\mathcal{G}} = \hat{\mathcal{G}}^{i,j} = \hat{\mathcal{G}}^{j,i}$.

On this supposition, first consider game $\hat{\mathcal{G}}^{1,2}$, which incorporates the LV-quarrel of $i=1$ with $j=2$. 
Let's calculate the {\sc yes}-successful sets $\hat{\mathcal{W}}^{1,2}$.
Observe that for any $j \in S$, $S \in \hat{\mathcal{W}}$ if and only if $S \setminus \{i\} \in \mathcal{W}$.
Thus $\{1,2\} \in \hat{\mathcal{W}}$ and $\{2\} \in \hat{\mathcal{W}}$ if and only if $\{2\} \in \mathcal{W}$.
But $\{2\} \notin \mathcal{W}$, so both $\{1,2\} \notin \hat{\mathcal{W}}$ and $\{2\} \notin \hat{\mathcal{W}}$.
Next, for any $j\notin S$, $S \in \hat{\mathcal{W}}$ if and only if $S \cup \{i\} \in \mathcal{W}$.
Thus $\emptyset \in \hat{\mathcal{W}}$ and $\{1\} \in \hat{\mathcal{W}}$ if and only if $\{1\} \in \mathcal{W}$.
But $\{1\} \in \mathcal{W}$, so both $\emptyset \in \hat{\mathcal{W}}$ and $\{1\} \in \hat{\mathcal{W}}$.
Hence $\hat{\mathcal{W}}^{1,2}=\{\emptyset, \{1\}\}$.

Again supposing the LV-rule satisfies reciprocality,
consider now $\hat{\mathcal{G}}^{2,1}$, which incorporates the LV-quarrel of $i=2$ with $j=1$.
Again, for any $j \in S$, $S \in \hat{\mathcal{W}}$ if and only if $S \setminus \{i\} \in \mathcal{W}$.
This implies $\{1,2\}\in \hat{\mathcal{W}}$ and $\{1\}\in \hat{\mathcal{W}}$ if and only if $\{1\} \in \mathcal{W}$.
But $\{1\} \in \mathcal{W}$, so both $\{1,2\}\in \hat{\mathcal{W}}$ and $\{1\}\in \hat{\mathcal{W}}$.
Next, for any $j\notin S$, $S \in \hat{\mathcal{W}}$ if and only if $S \cup \{i\} \in \mathcal{W}$.
Thus $\emptyset \in \hat{\mathcal{W}}$ and $\{2\} \in \hat{\mathcal{W}}$ if and only if $\{2\} \in \mathcal{W}$.
But $\{2\} \notin \mathcal{W}$, so $\emptyset \notin \hat{\mathcal{W}}$ and $\{2\} \notin \hat{\mathcal{W}}$.
It follows that $\hat{\mathcal{W}}^{2,1}=\{$\{1\}, \{1,2\}\}.

Recall, however, that $\hat{\mathcal{W}}^{1,2}=\{\emptyset, \{1\}\}$.
This implies that $\hat{\mathcal{G}}^{1,2}\neq \hat{\mathcal{G}}^{2,1}$,
which contradicts our initial supposition that an LV-quarrel is reciprocal.
\end{proof}

\restatetheorem{thm:LV-DNMQ}
The LV-rule is disposed to induce non-monotonicity over quarrellers.
\end{theorem}
\begin{proof}
Take any pair of quarrellers $\{i,j\}$ and any game $\mathcal{G}\in \mathbb{G}$ that satisfies condition $V(S)$ of DNMQ.
There thus exists an $S \ (i,j \notin S)$ where $S\cup \{i\} \in \mathcal{W}$ but $S\cup \{j\} \notin \mathcal{W}$.
Now consider an LV-quarrel $\mathcal{Q}$ where $i$ quarrels with $j$.
Since $S\cup \{j\} \notin \mathcal{W}$, by the LV-rule we have $S \cup \{i,j\} \notin \hat{\mathcal{W}}$.
Therefore, we have $S \cup \{i\} \in \hat{\mathcal{W}}$ and $S \cup \{i,j\} \notin \hat{\mathcal{W}}$ and so
$\mathcal{Q}$ is disposed to induce non-monotonicity over quarrellers.
\end{proof}

\subsection{Proofs for Section~\ref{sec:three-degrees}}

\restatetheorem{thm:a-r-w}
An asymmetric reciprocal weak {\sc yes}-quarrel is transformation monotonic.
\end{theorem}
\begin{proof}
The original game $\mathcal{G}$ is monotonic. 
But the only subsets that have changed from {\sc yes}-successful to {\sc yes}-unsuccessful are of the form $S \cup \{i,j\}$.
So suppose we now have a violation in monotonicity involving $S \cup \{i,j\}$. 
By definition, $S \cup \{i\} \notin \mathcal{W}$ and $S \cup \{j\} \notin \mathcal{W}$,
and hence $S \cup \{i\} \notin \hat{\mathcal{W}}$ and $S \cup \{j\} \notin \hat{\mathcal{W}}$.
So the violation must be due to a {\sc yes}-successful set $S \cup \{i,j\} \setminus k$.

Now consider $S\cup \{j\} \setminus k$ and 
$S\cup \{i\} \setminus k$.
Since both $S\cup \{i\} \notin \mathcal{W}$ and $S\cup \{j\} \notin \mathcal{W}$, by monotonicity 
we have $S \cup \{j\} \setminus \{k\} \notin \mathcal{W}$ and $S \cup \{i\} \setminus \{k\} \notin \mathcal{W}$.
But then, by definition, $S \cup \{i,j\} \setminus \{k\} \notin \hat{\mathcal{W}}$.
This contradicts the supposition that $S\cup \{i,j\}$ violates monotonicity in $\hat{\mathcal{G}}$.
\end{proof}

\restatetheorem{thm:a-r-s-DNMQ}
An asymmetric reciprocal strong {\sc yes}-quarrel $\mathcal{Q}$ is disposed to induce non-monotonicity over quarrellers.
\end{theorem}

\begin{proof}
Take any pair of quarrellers $\{i,j\}$ and any game $\mathcal{G}\in \mathbb{G}$ that satisfies condition $V(S)$ of DNMQ.
There thus exists an $S \ (i,j \notin S)$ where $S\cup \{i\} \in \mathcal{W}$ but $S\cup \{j\} \notin \mathcal{W}$.
Now consider a reciprocal strong {\sc yes}-quarrel $\mathcal{Q}$.
Since $\mathcal{G}$ is monotonic,  $S \notin \mathcal{W}$,
so by $\mathcal{Q}$ we have $S \cup \{i,j\} \notin \hat{\mathcal{W}}$.
Therefore, we have $S \cup \{i\} \in \hat{\mathcal{W}}$ and $S \cup \{i,j\} \notin \hat{\mathcal{W}}$ and so
$\mathcal{Q}$ is disposed to induce non-monotonicity over quarrellers.
\end{proof}

%

\restatetheorem{thm:s-r-w}
A symmetric reciprocal weak quarrel is transformation monotonic.
\end{theorem}
\begin{proof}
The original game $\mathcal{G}$ is monotonic. We now have two cases.
First, the only subsets that have changed from {\sc yes}-successful to {\sc yes}-unsuccessful are of the form $S\cup \{i,j\}$.
So suppose we have a violation in monotonicity involving $S \cup \{i,j\}$. 
By definition, $S \cup \{i\} \notin \mathcal{W}$ and $S \cup \{j\} \notin \mathcal{W}$,
and hence $S \cup \{i\} \notin \hat{\mathcal{W}}$ and $S \cup \{j\} \notin \hat{\mathcal{W}}$.
So the violation must be due to a {\sc yes}-successful set $S \cup \{i,j\} \setminus k$.

Now consider $S\cup \{j\} \setminus \{k\}$ and $S \cup \{i\} \setminus \{k\}$.
Since $S \cup \{i\} \notin \mathcal{W}$ and $S \cup \{j\} \notin \mathcal{W}$,
by monotonicity we have $S \cup \{j\} \setminus \{k\} \notin \mathcal{W}$ and $S \cup \{i\} \setminus \{k\} \notin \mathcal{W}$.
But then by definition, $S \cup \{i,j\} \setminus \{k\} \notin \hat{\mathcal{W}}$.
This contradicts the supposition that $S \cup \{i,j\}$ violates monotonicity in $\hat{\mathcal{G}}$.

Second, the only subsets $S$ that have changed from {\sc yes}-successful to {\sc yes}-unsuccessful are of the form where $i,j\notin S$.
So suppose we have a violation in monotonicity involving $S$.
By definition, $S \cup \{i\} \in \mathcal{W}$ and $S \cup \{j\} \in \mathcal{W}$ and hence $S \cup \{i\} \in \hat{\mathcal{W}}$ and $S \cup \{j\} \in \hat{\mathcal{W}}$.
So the violation must be due to a {\sc yes}-unsuccessful set $S \cup \{k\}$.

Now consider $S \cup \{j,k\}$ and $S \cup \{i,k\}$.
Since $S \cup \{i\} \in \mathcal{W}$ and $S \cup \{j\} \in \mathcal{W}$,
by monotonicity $S \cup \{j,k\} \in \mathcal{W}$ and $S \cup \{i,k\} \in \mathcal{W}$.
But then by definition, $S \cup \{k\} \in \hat{\mathcal{W}}$.
This contradicts the supposition that $S$ violates monotonicity in $\hat{\mathcal{G}}$.
\end{proof}

\restatetheorem{thm:s-r-s}
A symmetric reciprocal strong quarrel is disposed to induce non-monotonicity over quarrellers.
\end{theorem}
\begin{proof}
The proof is identical to that of Theorem \ref{thm:a-r-s-DNMQ}.
\end{proof}

\restatetheorem{thm:s-r-c}
A symmetric reciprocal cataclysmic quarrel is disposed to induce non-monotonicity over quarrellers.
\end{theorem}
\begin{proof}
The proof is identical to that of Theorem \ref{thm:FM-DNMQ}.
\end{proof}

\subsection{Proofs for Section~\ref{sec:nonrecip}}

\restatetheorem{thm:a-n-s-DNMQ}
An asymmetric non-reciprocal strong {\sc yes}-quarrel $\mathcal{Q}$ is disposed to induce non-monotonicity over quarrellers.
\end{theorem}
\begin{proof}
Take any pair of quarrellers $\{i,j\}$ and any game $\mathcal{G}\in \mathbb{G}$ that satisfies condition $V(S)$ of DNMQ.
There thus exists an $S \ (i,j \notin S)$ where $S\cup \{i\} \in \mathcal{W}$ but $S\cup \{j\} \notin \mathcal{W}$.
Now consider a non-reciprocal strong {\sc yes}-quarrel $\mathcal{Q}$ where $i$ quarrels with $j$.
Since $S\cup \{j\} \notin \mathcal{W}$, by $\mathcal{Q}$ we have $S \cup \{i,j\} \notin \hat{\mathcal{W}}$.
Therefore, we have $S \cup \{i\} \in \hat{\mathcal{W}}$ and $S \cup \{i,j\} \notin \hat{\mathcal{W}}$ and so
$\mathcal{Q}$ is disposed to induce non-monotonicity over quarrellers.
\end{proof}

\restatetheorem{thm:a-n-c-DNMQ}
An asymmetric non-reciprocal cataclysmic {\sc yes}-quarrel $\mathcal{Q}$ is disposed to induce non-monotonicity over quarrellers.
\end{theorem}

\begin{proof}
Take any pair of quarrellers $\{i,j\}$ and any game $\mathcal{G}\in \mathbb{G}$ that satisfies condition $V(S)$ of DNMQ.
There thus exists an $S \ (i,j \notin S)$ where $S\cup \{i\} \in \mathcal{W}$ but $S\cup \{j\} \notin \mathcal{W}$.
Now consider a non-repiprocal cataclysmic {\sc yes}-quarrel $\mathcal{Q}$ where $i$ quarrels with $j$.
Since $\mathcal{G}$ is monotonic, $\{j\} \notin \mathcal{W}$, and so by $\mathcal{Q}$ we have $S \cup \{i,j\} \notin \hat{\mathcal{W}}$.
Therefore, we have $S \cup \{i\} \in \hat{\mathcal{W}}$ and $S \cup \{i,j\} \notin \hat{\mathcal{W}}$ and so
$\mathcal{Q}$ is disposed to induce non-monotonicity over quarrellers.
\end{proof}

\restatetheorem{thm:s-n-c-DNMQ}
A symmetric non-reciprocal cataclysmic quarrel $\mathcal{Q}$ is disposed to induce non-monotonicity over quarrellers.
\end{theorem}

\begin{proof}
The proof is identical to that of Theorem \ref{thm:a-n-c-DNMQ}.
\end{proof}

\subsection{Proofs for Section~\ref{sec:conclusion}}

\restatetheorem{thm:SS}
The Shapley-Shubik Index satisfies the standard 
quarrel postulate based on a symmetric, weak quarrel.
\end{theorem}
\begin{proof}
Let's first recall the definition of voting power for the Shapley-Shubik Index.
Let $\sigma$ be an ordering (or permutation) of the $n$ players, and let $\Omega$ be the set of
all orderings. We say that player $i$ is {\em pivotal} for the ordering $\sigma=(\sigma_1,\sigma_2,\dots,\sigma_n)$ if
$\sigma_k=\{i\}$ and $i$ is {\sc yes}-decisive for $S=\{\sigma_1,\sigma_2,\dots,\sigma_k\}$.
We remark that for a monotonic game, there is a unique pivotal player for each ordering.
The a priori voting power of player $i$ under the Shapley-Shubik Index is then supposed to be:
$$\psi_i^{SS} \ =\ \sum_{\sigma\in \Omega} \mathbb{P}(\sigma)\cdot \chi_i(\sigma) \ = \ \sum_{\sigma\in \Omega} \frac{1}{n!}\cdot \chi_i(\sigma)$$
where
$$
\chi_i(\sigma) =
\begin{cases}
1& \mathrm{if\ } i \mathrm{\ is \ pivotal \ in\ } \sigma\\
0 & \mathrm{otherwise}
\end{cases}
$$
Now consider the game $\hat{\mathcal{G}}$ that incorporates a symmetric, weak quarrel between $i$ and $j$ in the original game $\mathcal{G}$. 
We must prove $\hat{\psi}^{SS}_i \le {\psi}^{SS}_i$. Take any ordering $\sigma=(\sigma_1,\sigma_2,\dots,\sigma_n)\in \Omega$.
Recall that $\hat{\mathcal{G}}$ is monotonic for a symmetric, weak quarrel.
Therefore, it suffices to prove that if $\sigma_k=\{i\}$ and $i$ is {\sc yes}-decisive for $S=\{\sigma_1,\sigma_2,\dots,\sigma_k\}$
in $\hat{\mathcal{G}}$ then it is also {\sc yes}-decisive in $\mathcal{G}$.
There are two cases:\\
(i) $j\in S$. By assumption, $S\setminus \{i\} \notin \hat{\mathcal{W}}$ and $S \in \hat{\mathcal{W}}$.
But $j\in S\setminus \{i\}$ so, by definition of a general quarrel, the non-ambush condition implies that
$S\setminus \{i\}\notin \mathcal{W}$. On the other hand, 
$\{i,j\}\subseteq S$. Thus the cooperative-success-reduction condition (CSR:YQ-1)
implies that $S\setminus \{i\}\in \mathcal{W}$. But this implies $i$ is {\sc yes}-decisive for $S$
in the original game $\mathcal{G}$, as desired.\\
(ii) $j\notin S$. By assumption, $S\setminus \{i\} \notin \hat{\mathcal{W}}$ and $S \in \hat{\mathcal{W}}$.
Now, since $i\in S$, the non-ambush condition implies that $S\in \mathcal{W}$.
On the other hand, $\{i,j\}\cap S\setminus \{i\} =\emptyset$. Thus the cooperative-success-reduction condition 
(namely, the contrapositive of CSR:NQ-1) implies that $S\setminus \{i\}\notin \mathcal{W}$. 
But this implies $i$ is {\sc yes}-decisive for $S$ in the original game $\mathcal{G}$, as desired.

If follows that the Shapley-Shubik Index satisfies the standard quarrel postulate.
\end{proof}

\restatetheorem{thm:PB}
The Penrose-Banzhaf Measure satisfies the standard 
quarrel postulate based on a symmetric, weak quarrel.
\end{theorem}
\begin{proof}
The a priori voting power of player $i$ under the Penrose-Banzhaf Measure can be given by:
$$\psi_i^{PB} \ =\ \sum_{S\subseteq N: i\in S} \frac{1}{2^{n-1}}\cdot \chi_i(S)$$
where
$$
\chi_i(S) =
\begin{cases}
1& \mathrm{if\ } i \mathrm{\ is \ {\text{{\sc yes}-decisive}} \ in\ } \mathbb{S}\\
0 & \mathrm{otherwise}
\end{cases}
$$
Now consider the game $\hat{\mathcal{G}}$ that incorporates a symmetric, weak quarrel between $i$ and $j$ 
in the original game $\mathcal{G}$.
Again, we must prove $\hat{\psi}^{PB}_i \le {\psi}^{PB}_i$. This follows by a similar argument as in the proof of
Theorem~\ref{thm:SS}.
\end{proof}

\end{document}